\documentclass[ALICE,manyauthors]{cernphprep}

\usepackage[comma,square,numbers,sort&compress]{natbib}
\usepackage{hyperref}
\usepackage{lineno}
\graphicspath{{../Results/Figures/}}
\usepackage{color}
\usepackage[normalem]{ulem}
\usepackage[inline]{enumitem}
\newcommand {\lint}      {{L_\mathrm{int}}}
\newcommand {\pp}        {\ensuremath{\mathrm {p\kern-0.05em p}}}
\newcommand {\PbPb}      {\ensuremath{\mathrm{Pb\mbox{--}Pb}}}

\newcommand {\mmumu}      {{m_{\mu\mu}}}

\newcommand {\ccbar}     {\ensuremath{c\bar{c}}}

\newcommand {\jpsi}      {\ensuremath{\mathrm{J}\kern-0.02em/\kern-0.05em\psi}}
\newcommand {\psip}      {\ensuremath{\psi\mathrm{(2S)}}}

\newcommand {\pt}        {\ensuremath{p_{\mathrm{\textsc{t}}}}}

\newcommand {\meanpt}    {\ensuremath{\langle p_{\mathrm{\textsc{t}}} \rangle}}
\newcommand {\meanptsq}  {\ensuremath{\langle p_{\mathrm{\textsc{t}}}^{2} \rangle}}
\newcommand {\pttrig}    {\ensuremath{p_{\mathrm{\textsc{t}}}^{\mathrm{trig}}}}
\newcommand {\y}         {\ensuremath{y}}

\newcommand {\sqrts}             {\ensuremath{\sqrt{s}}\,}

\newcommand {\br}        {{\mathrm{BR}}}

\newcommand {\mevc}      {\ensuremath{\,\mathrm{MeV}\kern-0.05em/\kern-0.02em c}}
\newcommand {\mevcc}     {\ensuremath{\,\mathrm{MeV}\kern-0.05em/\kern-0.02em c^2}}
\newcommand {\gevc}      {\ensuremath{\,\mathrm{GeV}\kern-0.05em/\kern-0.02em c}}
\newcommand {\gevcc}     {\ensuremath{\,\mathrm{GeV}\kern-0.05em/\kern-0.02em c^2}}
\newcommand {\gevcsq}    {\ensuremath{\,\mathrm{GeV}^2\kern-0.05em/\kern-0.02em c^2}}
\newcommand {\gevfmcube} {\ensuremath{\,\mathrm{GeV}\kern-0.05em/\kern-0.02em \mathrm{fm}^3}}
\newcommand {\Ae}        {\ensuremath{A\varepsilon}}

\renewcommand{\arraystretch}{1.2} % more vertical space in tables, look nicer

\newcommand{\dif}{{\textup d}}

\begin{document}%

%%%%%%%%%%%%%%%  Title page %%%%%%%%%%%%%%%%%%%%%%%%
\begin{titlepage}
\PHyear{2017}
\PHnumber{015}      % required, will be obtained from PH
\PHdate{25 January}  % required, will be obtained from PH
%

%%% Put your own title + short title here:
\title{Energy dependence of forward-rapidity $\jpsi$ and $\psip$ production \\in pp collisions at the LHC}
\ShortTitle{Energy dependence of $\jpsi$ and $\psip$ production in pp collisions at the LHC}   % appears on right page headers

%%% Do not change the next lines
\Collaboration{ALICE Collaboration\thanks{See Appendix~\ref{app:collab} for the list of collaboration members}}
\ShortAuthor{ALICE Collaboration} % appears on left page headers, do not change

\begin{abstract}
We present results on transverse momentum ($\pt$) and rapidity ($\y$) differential production cross sections, mean transverse momentum and mean transverse momentum square of inclusive $\jpsi$ and $\psip$ at forward rapidity ($2.5<y<4$) as well as $\psip$-to-$\jpsi$ cross section ratios.These quantities are measured in pp collisions at center of mass energies $\sqrts=5.02$ and $13$~TeV with the ALICE detector. Both charmonium states are reconstructed in the dimuon decay channel, using the muon spectrometer. A comprehensive comparison to inclusive charmonium cross sections measured at $\sqrts=2.76$, $7$ and $8$~TeV is performed. A comparison to non-relativistic quantum chromodynamics and fixed-order next-to-leading logarithm calculations, which describe prompt and non-prompt charmonium production respectively, is also presented. A good description of the data is obtained over the full $\pt$ range, provided that both contributions are summed. In particular, it is found that for $\pt>15$~GeV/$c$ the non-prompt contribution reaches up to 50\% of the total charmonium yield.

\end{abstract}
\end{titlepage}
\setcounter{page}{2}

\section{Introduction}
Charmonia, such as $\jpsi$ and $\psip$, are bound states of a charm and anti-charm quark ($\ccbar$). 
At LHC energies, their hadronic production results mostly from the hard scattering of two gluons into a $\ccbar$ pair followed by the evolution of this pair into a charmonium state. Charmonium measurements in pp collisions are essential to the investigation of their production mechanisms. They also provide a baseline for proton-nucleus and nucleus-nucleus results which in turn are used to quantify the properties of the quark-gluon plasma~\cite{Brambilla:2010cs,Andronic:2015wma}. 

Mainly three theoretical approaches are used to describe the hadronic production of charmonium: the Color Evaporation Model (CEM)~\cite{Fritzsch:1977ay,Amundson:1996qr}, the Color Singlet Model (CSM)~\cite{Baier:1981uk} and the Non-Relativistic Quantum Chro\-mo-Dy\-na\-mics model (NRQCD)~\cite{Bodwin:1994jh}. These approaches differ mainly in the treatment of the evolution of the heavy-quark pair into a bound state. In the CEM, the production cross section of a given charmonium is proportional to the $\ccbar$ cross section, integrated between the mass of the charmonium and twice the mass of the lightest D meson, with the proportionality factor being independent of the charmonium transverse momentum $\pt$, rapidity $y$ and of the collision center of mass energy $\sqrts$.
In the CSM, perturbative QCD is used to describe the $\ccbar$ production with the same quantum numbers as the final-state meson. In particular, only color-singlet (CS) $\ccbar$ pairs are considered. Finally, in the NRQCD framework charmonium can be formed from a $\ccbar$ pair produced either in a CS or in a color-octet (CO) state. The color neutralization of the CO state is treated as a non-perturbative process. For a given order in $\alpha_s$, it is expanded in powers of the relative velocity between the two charm quarks and parametrized using universal Long Distance Matrix Elements (LDME) which are fitted to the data. The predictive power of NRQCD calculations is tested by fitting the LDME to a subset of the data and comparing cross sections calculated with these LDME to measurements performed at different energies.
It is therefore crucial to confront these models to as many measurements as possible, over a wide range of $\pt$, $y$ and $\sqrts$, and with as many different charmonium states as possible. The comparison can also be extended to observables other than cross sections, such as charmonium polarization~\cite{alice:2012pol,Aaij:2013nlm,Aaij:2014qea}.

In this paper we present results on the production cross sections of inclusive $\jpsi$ and $\psip$ at forward rapidity ($2.5<y<4$) measured in pp collisions at center of mass energies $\sqrts=13$ and $5.02$~TeV. For $\jpsi$ at $\sqrts=5.02$~TeV, the $\pt$-differential cross sections have been published in~\cite{Adam:2016rdg} while the $y$-differential cross sections are presented here for the first time.

The $\jpsi$ and $\psip$ are measured in the dimuon decay channel. The inclusive differential cross sections are obtained as a function of $\pt$ and $\y$ over the ranges $0<\pt<30$~GeV/$c$ for $\jpsi$ at $\sqrts=13$~TeV, $0<\pt<12$~GeV/$c$ for $\jpsi$ at $\sqrts=5.02$~TeV and $0<\pt<16$~GeV/$c$ for $\psip$ at $\sqrts=13$~TeV. 
At $\sqrts=5.02$~TeV only the $\pt$-integrated $\psip$ cross section is measured due to the limited integrated luminosity.
The $\jpsi$ result at $\sqrts=13$~TeV extends significantly the $\pt$ reach of measurements performed in a similar rapidity range by LHCb~\cite{Aaij:2015rla}. The $\jpsi$ result at $\sqrts=5.02$~TeV and the $\psip$ results at both $\sqrts$ are the first at this rapidity. The inclusive $\psip$-to-$\jpsi$ cross section ratios as a function of both $\pt$ and $y$ are also presented. These results are compared to similar measurements performed at $\sqrts=2.76$~\cite{Abelev:2012kr}, $7$~\cite{Abelev:2014qha} and $8$~TeV~\cite{Adam:2015rta}. These comparisons allow studying the variations of quantities such as the mean transverse momentum $\meanpt$, mean transverse momentum square $\meanptsq$ and the $\pt$-integrated cross section as a function of $\sqrts$. Put together, these measurements constitute a stringent test for models of charmonium production. In particular, an extensive comparison of the $\jpsi$ and $\psip$ cross sections at all available collision energies to the calculations from two NRQCD groups is presented towards the end of the paper (Sec.~\ref{section:results}). In addition, the $\pt$-integrated $\jpsi$ cross section as a function of $\sqrts$ is also compared to a CEM calculation. No comparison to the CSM is performed since complete calculations are not available at these energies beside the ones published in~\cite{Abelev:2014qha, Lansberg:2011hi}.

All cross sections reported in this paper are inclusive and contain, on top of the direct production of the charmonium, a contribution from the decay of heavier charmonium states as well as contributions from the decay of long-lived beauty flavored hadrons ($b$-hadrons). The first two contributions (direct production and decay from heavier charmonium states) are commonly called prompt, whereas the contribution from $b$-hadron decays is called non-prompt because of the large mean proper decay length of these hadrons ($\sim$ $500$~$\mu$m).

The paper is organized as follows: the ALICE apparatus and the data samples used for this analysis are described in Sec.~\ref{sec:detector}, the analysis procedure is discussed in Sec.~\ref{sec:analysis} while the results are presented and compared to measurements at different $\sqrts$ as well as to models in Sec.~\ref{section:results}.

%__________________________________________________________
\section{Apparatus and data samples}\label{sec:detector}

The ALICE detector is described in detail in~\cite{Aamodt:2008zz,Abelev:2014ffa}. In this section, we introduce the detector subsystems relevant to the present analysis: the muon spectrometer, the Silicon Pixel Detector (SPD), the V0 scintillator hodoscopes and the T0 Cherenkov detectors. 

The muon spectrometer~\cite{Aamodt:2011gj} allows the detection and characterization of muons in the pseudorapidity range $-4<\eta<-2.5$~\footnote{We note that the ALICE reference frame defines the positive $z$ direction along the counter-clockwise beam direction, resulting in a negative pseudorapidity range for detectors like the muon spectrometer. However, due to the symmetry of pp collisions, the rapidity is kept positive when presenting results.}. It consists of a ten-interaction-lengths front absorber followed by a 3 T$\cdot$m dipole magnet coupled to a system of tracking (MCH) and triggering (MTR) devices. The front absorber is placed between 0.9 and 5 m from the Interaction Point (IP) and filters out hadrons and low-momentum muons emitted at forward rapidity. Tracking in the MCH is performed using 5 stations, each one consisting of two planes of cathode pad chambers positioned between 5.2 and 14.4 m from the IP. The MTR is positioned downstream of a 1.2 m thick iron wall which absorbs the remaining hadrons that escape the front absorber as well as low-momentum muons. It is composed of two stations equipped with two planes of resistive plate chambers each placed at 16.1 and 17.1 m from the IP. A conical absorber ($\theta<2^\circ$) protects the muon spectrometer against secondary particles produced mainly by large-$\eta$ primary particles interacting with the beam pipe throughout its full length. Finally, a rear absorber located downstream of the spectrometer protects the MTR from the background generated by beam-gas interactions.

The SPD is used to reconstruct the primary vertex of the collision. It is a cylindrically-shaped silicon pixel tracker and corresponds to the two innermost layers of the Inner Tracking System (ITS)~\cite{Aamodt:2010aa}. These two layers surround the beam pipe at average radii of 3.9 and 7.6 cm and cover the pseudorapidity intervals $|\eta|<2$ and $|\eta|<1.4$, respectively.

The V0 hodoscopes~\cite{Abbas:2013taa} consist of two scintillator arrays positioned on each side of the IP at $z=-90$ and $340$~cm and covering the $\eta$ range $-3.7<\eta<-1.7$ and $2.8<\eta<5.1$ respectively. They are used for online triggering and to reject beam-gas events by means of offline timing cuts together with the T0 detectors.

Finally, the T0 detectors~\cite{Bondila:2005xy} are used for the luminosity determination. They consist of two arrays of quartz Cherenkov counters placed on both sides of the IP covering the $\eta$ ranges $-3.3<\eta<-3$ and $4.6<\eta<4.9$.

The data used for this paper were collected in 2015. They correspond to pp collisions at $\sqrts=13$ and $5.02$~TeV.
The data at $\sqrts=13$~TeV are divided into several sub-periods corresponding to different beam conditions and leading to different pile-up rates. The pile-up rate, defined as the probability that one recorded event contains two or more collisions, reaches up to 25\% in the muon spectrometer for beams with the highest luminosity. The data at $\sqrts=5.02$~TeV were collected during the 5 days immediately after the $\sqrts=13$~TeV campaign. During this period the pile-up rate was stable and below 2.5\%. 

Events used for this analysis were collected using a dimuon trigger which requires that two muons of opposite sign are detected in the MTR in coincidence with the detection of a signal in each side of the V0. In addition, the transverse momentum $\pttrig$ of each muon, evaluated online, is required to pass a threshold of 0.5~GeV/$c$ (1~GeV/$c$) for the data taking at $\sqrts=5.02$ ($13$)~TeV in order to reject soft muons from $\pi$ and K decays and to limit the trigger rate when the instantaneous luminosity is high. This threshold is defined as the $\pt$ value for which the single muon trigger efficiency reaches 50\%~\cite{mtrBossu2012}.

The data samples available after the event selection described above correspond to an integrated luminosity $\lint=3.19\pm0.11$~pb$^{-1}$  and $\lint=106.3\pm2.2$~nb$^{-1}$ for $\sqrts=13$~TeV and $\sqrts=5.02$~TeV respectively. These integrated luminosities are measured following the procedure described in~\cite{ALICE-PUBLIC-2016-002} for the data at $\sqrts=13$~TeV and in~\cite{ALICE-PUBLIC-2016-005} for those at $\sqrts=5.02$~TeV. The systematic uncertainty on these quantities contains contributions from the measurement of the T0 trigger cross section using the Van der Meer scan technique~\cite{vanderMeer:1968zz} and the stability of the T0 trigger during data taking. The quadratic sum of these contributions amounts to $3.4$\% at $\sqrts=13$~TeV and $2.1$\% at $\sqrts=5.02$~TeV. 

\section{Analysis}\label{sec:analysis}

The differential production cross section for a charmonium state $\psi$ in a given $\pt$ and $y$ interval is:

\begin{equation}
	\frac{\dif^2\sigma_{\psi}}{\dif\pt\dif y}= \frac{1}{\Delta\pt\Delta y}\frac{1}{L_{\rm int}}\frac{N_{\psi}(\pt,\y)}{{\br}_{\psi\rightarrow\mu^+\mu^-}\Ae(\pt,\y)},
	\label{eq_production_cross_section}
\end{equation}

where $\br_{\psi\rightarrow\mu^+\mu^-} $ is the branching ratio of the charmonium state $\psi$ into a pair of muons ($5.96\pm0.03$\% for $\jpsi$ and $0.79\pm0.09$\% for $\psip$) ~\cite{Olive:2016xmw}, $\Delta\pt$ and $\Delta y$ are the widths of the $\pt$ and $y$ interval under consideration, $N_{\psi}(\pt,y)$ is the number of charmonia measured in this interval, $\Ae(\pt,\y)$ are the corresponding acceptance and efficiency corrections and $\lint$ is the integrated luminosity of the data sample. The large pile-up rates mentioned in Sec.~\ref{sec:detector} for the $\sqrts=13$~TeV data sample are accounted for in the calculation of $\lint$~\cite{ALICE-PUBLIC-2016-002}.

\subsection{Track selection}

The number of charmonia in a given $\pt$ and $y$ interval is obtained by forming pairs of opposite-sign muon tracks detected in the muon spectrometer and by calculating the invariant mass of these pairs, $\mmumu$. The resulting distribution is then fitted with several functions that account for both the charmonium signal and the background.

The procedure used to reconstruct muon candidates in the muon spectrometer is described in~\cite{Aamodt:2011gj}. Once muon candidates are reconstructed, additional offline criteria are applied in order to improve the quality of the dimuon sample and the signal-to-background (S/B) ratio. 

Tracks reconstructed in the MCH are required to match a track in the MTR which satisfies the single muon trigger condition mentioned in Sec.~\ref{sec:detector}. 
Each muon candidate is required to have a pseudorapidity in the interval $-4<\eta<-2.5$ in order to match the acceptance of the muon spectrometer. Finally, a cut on the transverse coordinate of the muon ($R_{\rm abs}$) measured at the end of the front absorber, $17.5<R_{abs}<89$~cm, ensures that muons emitted at small angles and passing through the high density section of the front absorber are rejected. 

These selection criteria remove most of the background tracks consisting of hadrons escaping from or produced in the front absorber, low-$\pt$ muons from $\pi$ and K decays, secondary muons produced in the front absorber and fake tracks. They improve the S/B ratio by up to 30\% for the $\jpsi$ and by a factor 2 for $\psip$.

\subsection{Signal extraction}
In each dimuon $\pt$ and $y$ interval, several fits to the invariant mass distribution are performed over different invariant mass ranges and using various fitting functions in order to obtain the number of $\jpsi$ and $\psip$ and to evaluate the corresponding systematic uncertainty. In all cases, the fit function consists of a background to which two signal functions are added, one for the $\jpsi$ and one for the $\psip$.

\begin{figure}[t!]
\centering
\begin{tabular}{cc}
\includegraphics[width=0.48\textwidth]{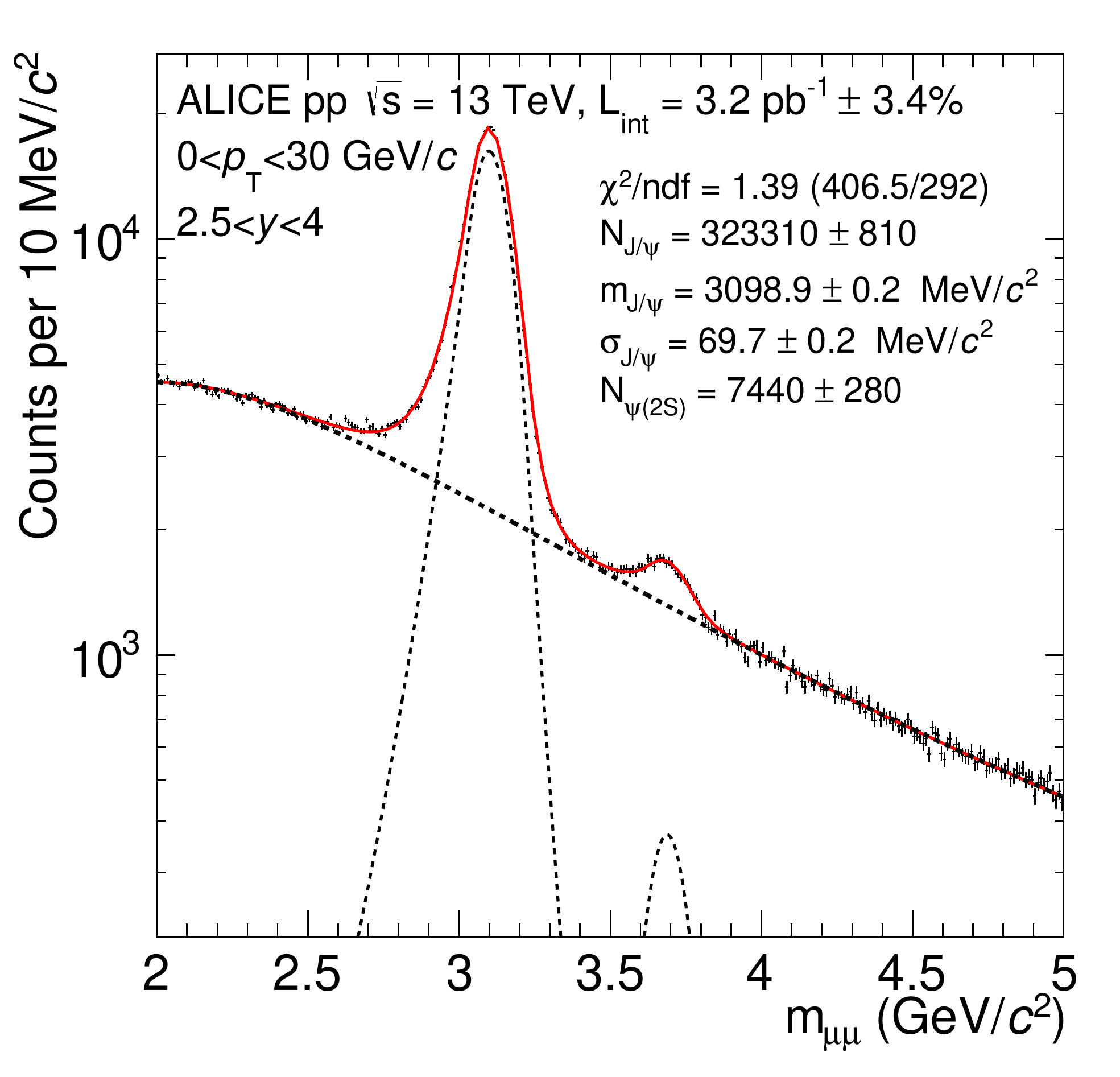}&
\includegraphics[width=0.48\textwidth]{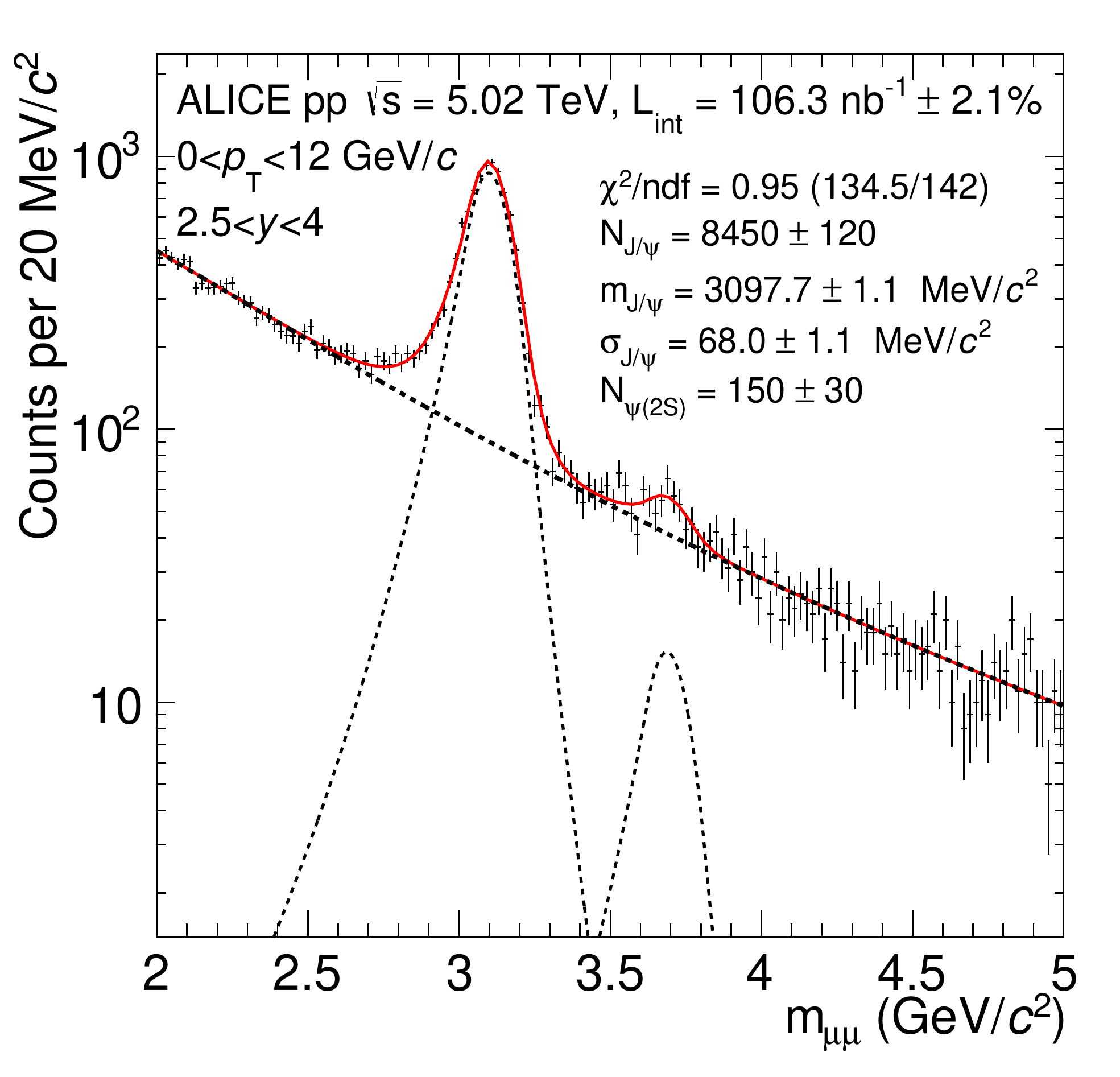}
\end{tabular}
\caption{Example of fit to the opposite-sign dimuon invariant mass distributions in pp collisions at $\sqrts=13$~TeV (left) and $5.02$~TeV (right). Dashed lines correspond to either signal or background functions, whereas the solid line corresponds to the sum of the signal and background functions.}
	\label{Fit}
\end{figure}

At $\sqrts=13$~TeV, the fits are performed over the invariant mass ranges $2.2<\mmumu<4.5$~GeV/$c^2$ and $2<\mmumu<5$~GeV/$c^2$. The background is described by either a pseudo-Gaussian function whose width varies linearly with the invariant mass or the product of a fourth-order polynomial and an exponential form.  The $\jpsi$ and $\psip$ signals are described by the sum of either two Crystal Ball or two pseudo-Gaussian functions~\cite{ALICE-PUBLIC-2015-006}.
These two signal functions consist of a Gaussian core with tails added on the sides that fall off slower than a Gaussian function. 
In most $\pt$ and $y$ intervals the parameters entering the definition of these tails cannot be left free in the fit due to the poor S/B ratio in the corresponding invariant mass region. They are instead fixed either to the values obtained from Monte Carlo (MC) simulations described in Sec.~\ref{Sec:accEff}, or to those obtained when fitting the measured $\pt$- and $y$-integrated invariant mass distribution with these parameters left free. For the $\jpsi$, the position, width and normalization of the signal are free parameters of the fit. For the $\psip$ only the normalization is free, whereas the position and width are bound to those of the $\jpsi$ following the same procedure as in~\cite{Adam:2015rta}. Finally, in all fits the background parameters are left free.

An identical approach is used at $\sqrts=5.02$~TeV, albeit with different invariant mass fitting ranges ($1.7<\mmumu<4.8$~GeV/$c^2$ and $2<\mmumu<4.4$~GeV/$c^2$) and a different set of background functions (a pseudo-Gaussian function or the ratio between a first- and a second-order polynomial function). For the signal the tails parameters are either fixed to those obtained in MC or taken from the $\sqrts=13$~TeV analysis.

The number of charmonia measured in a given $\pt$ and $y$ interval and the corresponding statistical uncertainty are taken as the mean of the values and uncertainties obtained from all the fits performed in this interval. The root mean square of these values is used as a systematic uncertainty.

Examples of fits to the $\pt$- and $y$-integrated invariant mass distributions are shown in Fig.~\ref{Fit}, at $\sqrts=13$ (left) and $5.02$~TeV (right). About $331\times10^3$ $\jpsi$ and $8.1\times10^3$ $\psip$ are measured at $\sqrts=13$~TeV whereas about $8.6\times10^3$ $\jpsi$ and $160$~$\psip$ are measured at $\sqrts=5.02$~TeV. Corresponding S/B ratios, evaluated within three standard deviations with respect to the charmonium pole mass, are 3.4 (4.5) for $\jpsi$ and 0.15 (0.18) for $\psip$ at $\sqrts=13$ ($5.02$)~TeV.

\subsection{Acceptance and efficiency corrections}\label{Sec:accEff}

Acceptance and efficiency corrections are obtained using MC simulations by computing the ratio between the number of charmonia reconstructed in the muon spectrometer and the number of generated charmonia in the same $\pt$ and $y$ interval. Independent simulations are performed for $\jpsi$ and $\psip$ and for each collision energy. Charmonia are generated using input $\pt$ and $y$ distributions obtained iteratively from the data. They are decayed into two muons using EVTGEN~\cite{Lange:2001uf} and PHOTOS~\cite{Barberio:1990ms} to properly account for the possible emission of accompanying radiative photons. 
It is assumed that both $\jpsi$ and $\psip$ are unpolarized consistently with the small longitudinal values reported in~\cite{alice:2012pol,Aaij:2013nlm,Aaij:2014qea} and accounting for further dilution coming from non-prompt charmonia.
The decay muons are tracked through a GEANT3~\cite{GEANT} model of the apparatus that includes a realistic description of the detectors and their performance during data taking. Track reconstruction and signal extraction are performed from the simulated hits generated in the detector using the same procedure and selection criteria as those used for the data.

The systematic uncertainty on acceptance and efficiency corrections contains the following contributions:
\begin{enumerate*}[label=(\roman*)]
\item the parametrization of the input $\pt$ and $\y$ distributions, 
%and the correlation between these,
\item the uncertainty on the tracking efficiency in the MCH,
\item the uncertainty on the MTR efficiency and 
\item the matching between tracks reconstructed in the MCH and tracks in the MTR.
\end{enumerate*}

For the parametrization of the MC input distributions, two sources of systematic uncertainty are considered: the correlations between $\pt$ and $y$ (more explicitely, the fact that the $\pt$ distribution of a given charmonium state varies with the rapidity interval in which it is measured~\cite{Aaij:2015rla}) and the effect of finite statistics in the data used to parametrize these distributions. At $\sqrts=5.02~$TeV, both contributions are evaluated by varying the input $\pt$ and $y$ distributions within limits that correspond to these effects and re-calculating the $\Ae$ corrections in each case as done in~\cite{Abelev:2014qha}. This corresponds to a variation of the input yields of at most 15\% as a function of $y$ and 50\% as a function of $\pt$. For $\jpsi$ measurements at $\sqrts=13$~TeV a slightly different approach is adopted in order to further reduce the sensitivity of the simulations to the input $\pt$ and $y$ distributions. It consists in evaluating the acceptance and efficiency corrections in small 2-dimensional bins of $y$ and $\pt$. These corrections are then applied on a dimuon pair-by-pair basis when forming the invariant mass distribution rather than applying them on the total number of measured charmonia in a given (larger) $\pt$ and $y$ interval. For each pair the corrections that match its $\pt$ and $y$ are used, thus making the resulting $\Ae$-corrected invariant mass distribution largely independent from the $\pt$ and $y$ distributions used as input to the simulations. For $\psip$ this improved procedure is not applied because the uncertainties on the measurement are dominated by statistics and the same method as for $\jpsi$ at $\sqrts=5.02$~TeV is used instead.

The other three sources of systematic uncertainty (tracking efficiency in the MCH, MTR efficiency, and matching between MTR and MCH tracks) are evaluated using the same procedure as in~\cite{Abelev:2014qha}, by comparing data and MC at the single muon level and propagating the observed differences to the dimuon case.

\subsection{Summary of the systematic uncertainties}
Table~\ref{table:systematics} gives a summary of the relative systematic uncertainties on the charmonium cross sections measured at $\sqrts=13$ and $\sqrts=5.02$~TeV. The total systematic uncertainty is the quadratic sum of all the sources listed in this table.
The uncertainty on the branching ratio is fully correlated between all measurements of a given state. 
The uncertainty on the integrated luminosity is fully correlated between measurements performed at the same $\sqrts$ and considered as uncorrelated from one $\sqrts$ to the other. The uncertainty on the signal extraction is considered as uncorrelated as a function of $\pt$, $y$ and $\sqrts$, but partially correlated between $\jpsi$ and $\psip$. Finally, all other sources of uncertainty are considered as partially correlated across measurements at the same energy and uncorrelated from one energy to the other. 

The systematic uncertainties on the MTR and MCH efficiencies are significantly smaller for the data at $\sqrts=5.02$~TeV than at $\sqrts=13$~TeV. This is due to the fact that the corresponding data taking period being very short, the detector conditions were more stable and therefore simpler to describe in the simulation. 
\begin{table}[h]
\centering
\begin{tabular}{c|cc|cc}
\hline
&\multicolumn{2}{c|}{$\sqrts=13$~TeV}&\multicolumn{2}{c}{$\sqrts=5.02$~TeV}\\
\hline
Source            &$\jpsi$ (\%)   &$\psip$ (\%)           &$\jpsi$ (\%)&$\psip$ (\%)\\
\hline
Branching ratio  &$0.6$            &$11$            &$0.6$          &$11$  \\
Luminosity       &$3.4$            &$3.4$           &$2.1$          &$2.1$ \\
Signal extraction&$3$ $(3-8)$      & $5$ $(5-9)$    &$3$ $(1.5-10)$ &$8$   \\
MC input         &$0.5$ $(0.5-1.5)$&$1$ $(0.5-4)$   &$2$ $(0.5-2.5)$&$2.5$ \\
MCH efficiency   &$4$              &$4$             &$1$            &$1$   \\
MTR efficiency   &$4$ $(1.5-4)$    &$4$ $(1.5-4)$   &$2$ $(1.5-2)$  &$2$   \\
Matching         &$1$              &$1$             &$1$            &$1$   \\
\hline
\end{tabular}
\caption{\label{table:systematics}Relative systematic uncertainties associated to the $\jpsi$ and $\psip$ cross section measurements at $\sqrts=13$ and $5.02$~TeV. Values in parenthesis correspond to the minimum and maximum values as a function of $\pt$ and $y$. For $\psip$ at $\sqrts=5.02$~TeV, only the $\pt$-integrated values are reported.}
\end{table}
\section{Results}\label{section:results}

\subsection{\label{sec:results13TeV}Cross sections and cross section ratios at $\sqrts=~13$ and $5.02$~TeV}
\begin{figure}[t!]
\centering
\begin{tabular}{cc}
\includegraphics[width=0.48\textwidth]{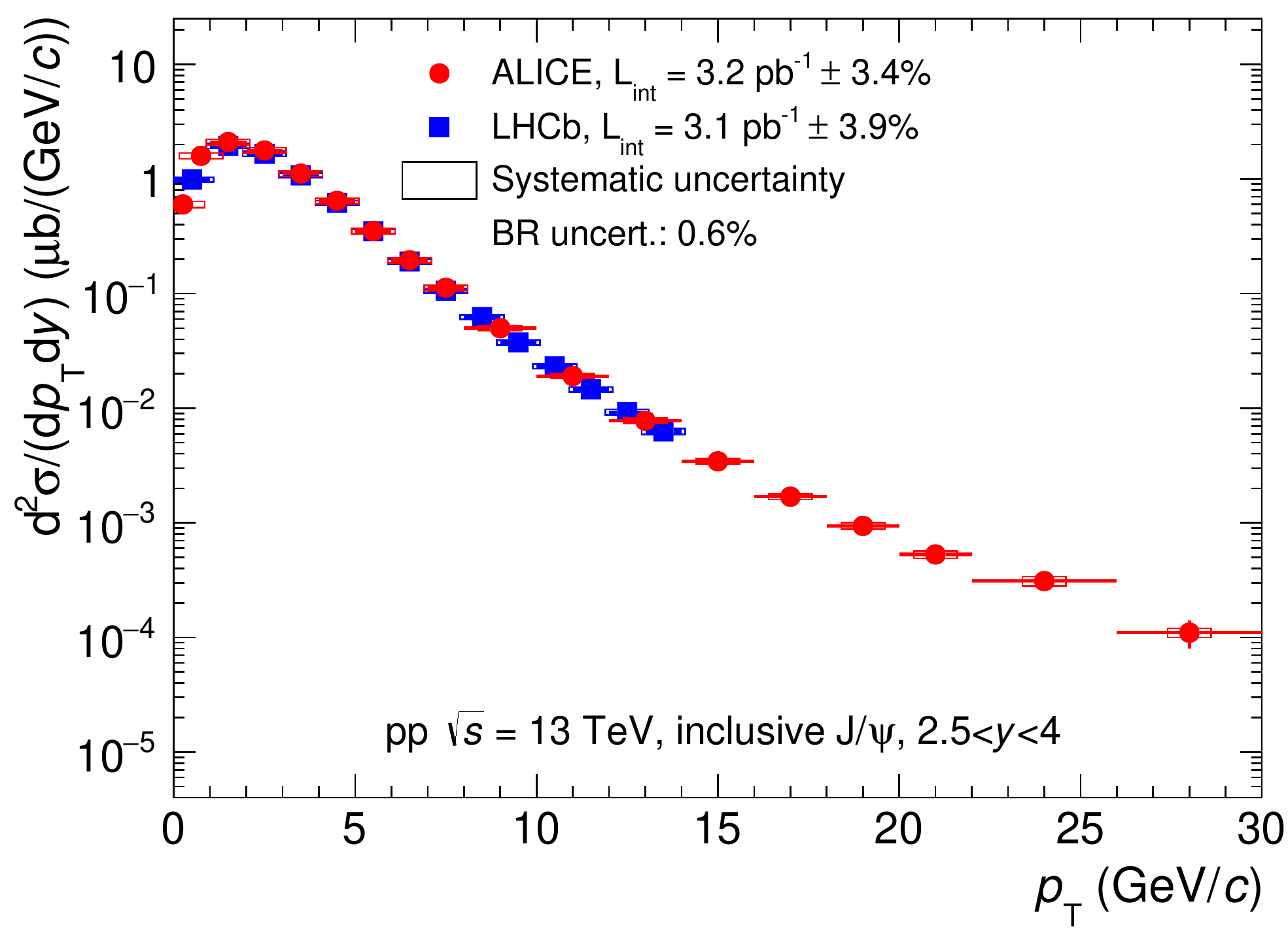}&\includegraphics[width=0.48\textwidth]{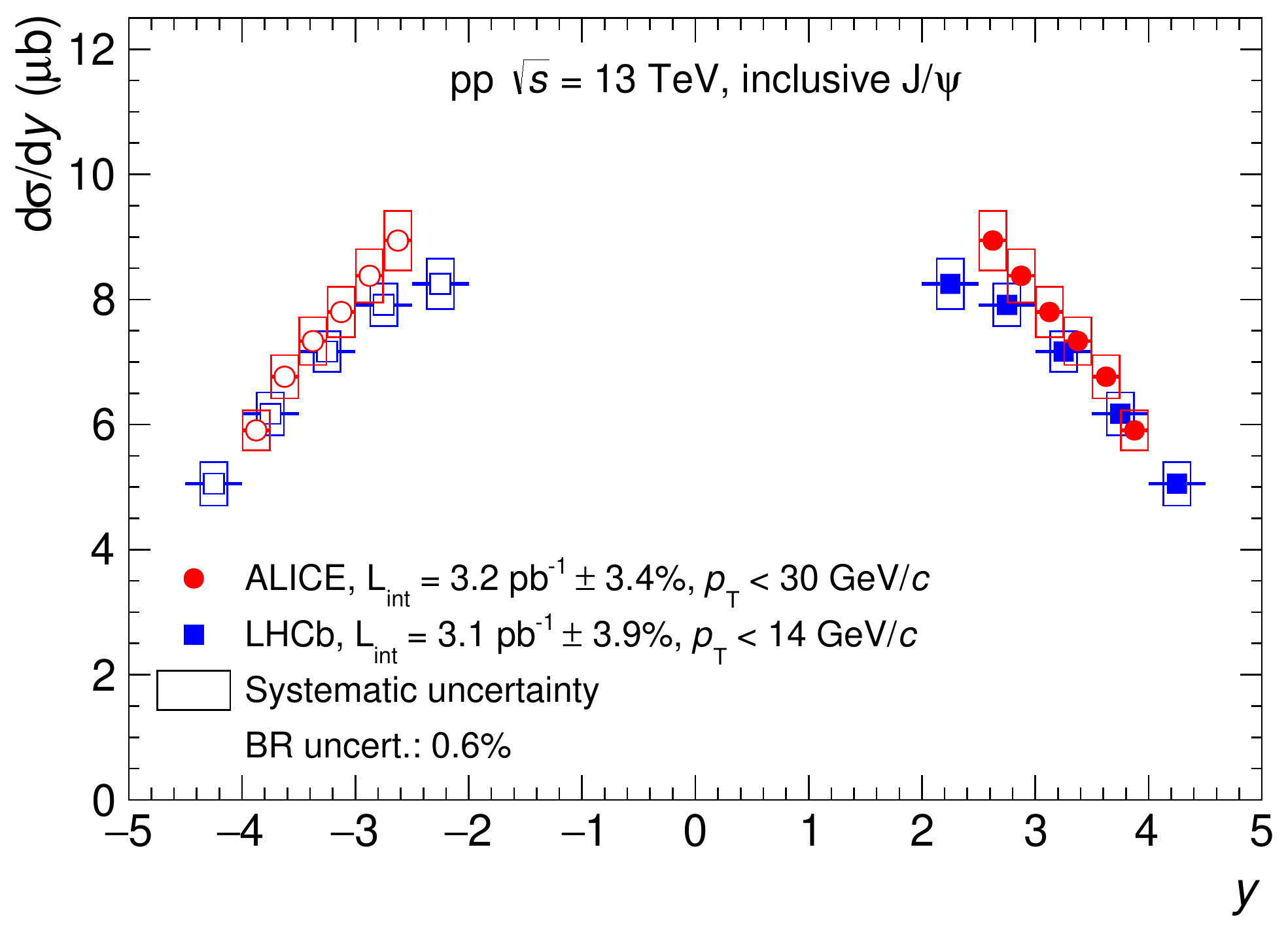}\\
\includegraphics[width=0.48\textwidth]{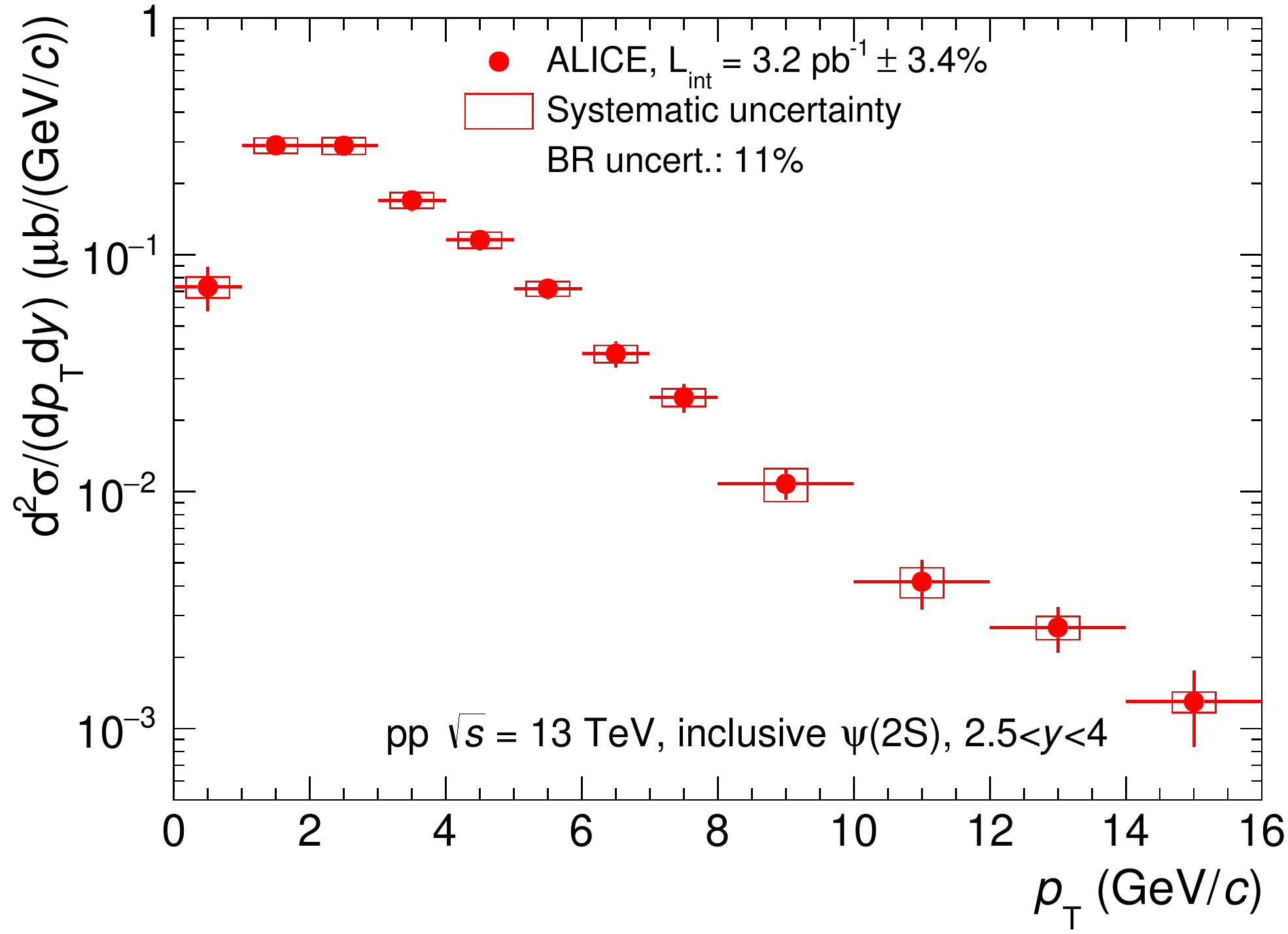}&\includegraphics[width=0.48\textwidth]{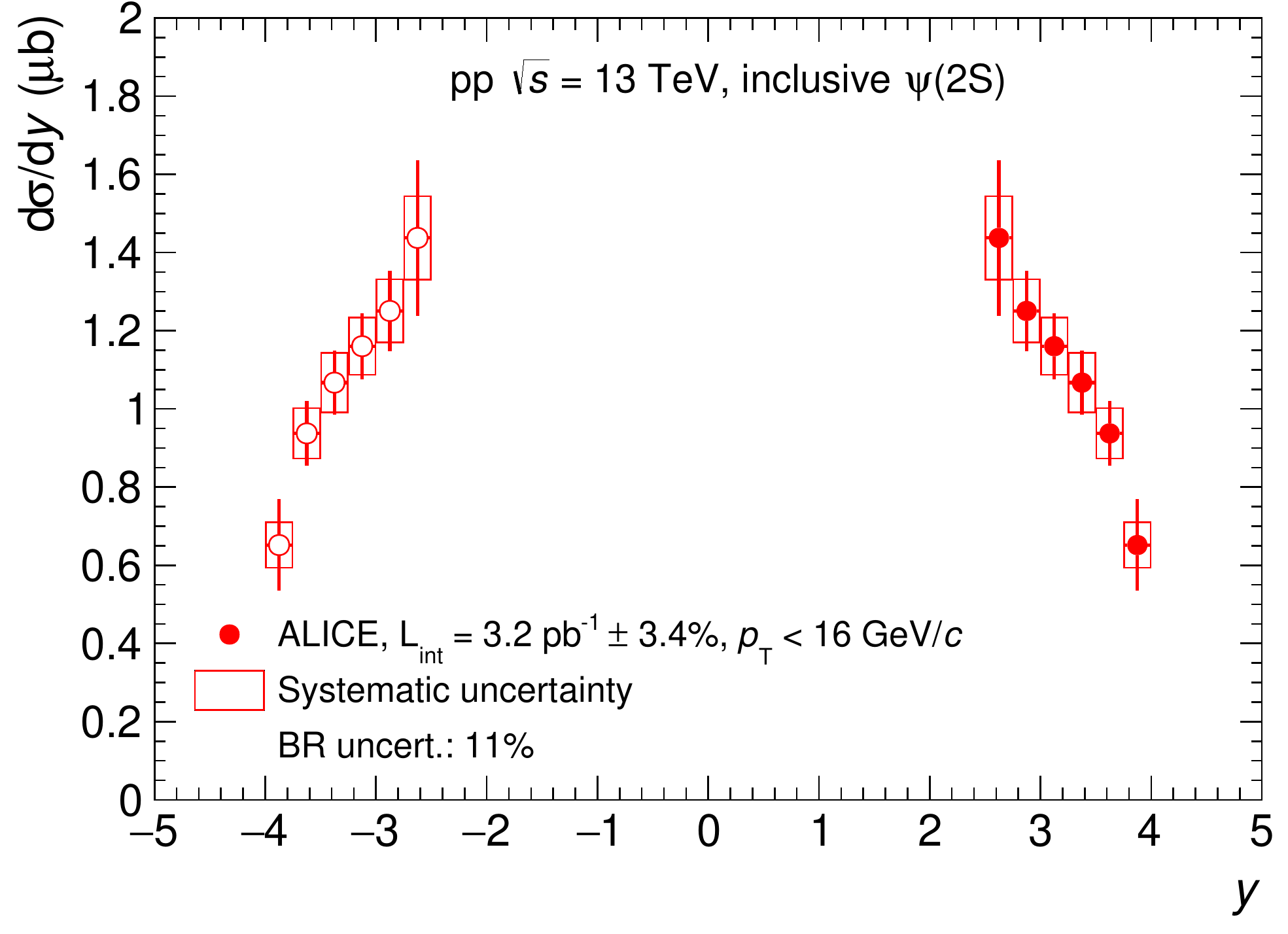}\\
\includegraphics[width=0.48\textwidth]{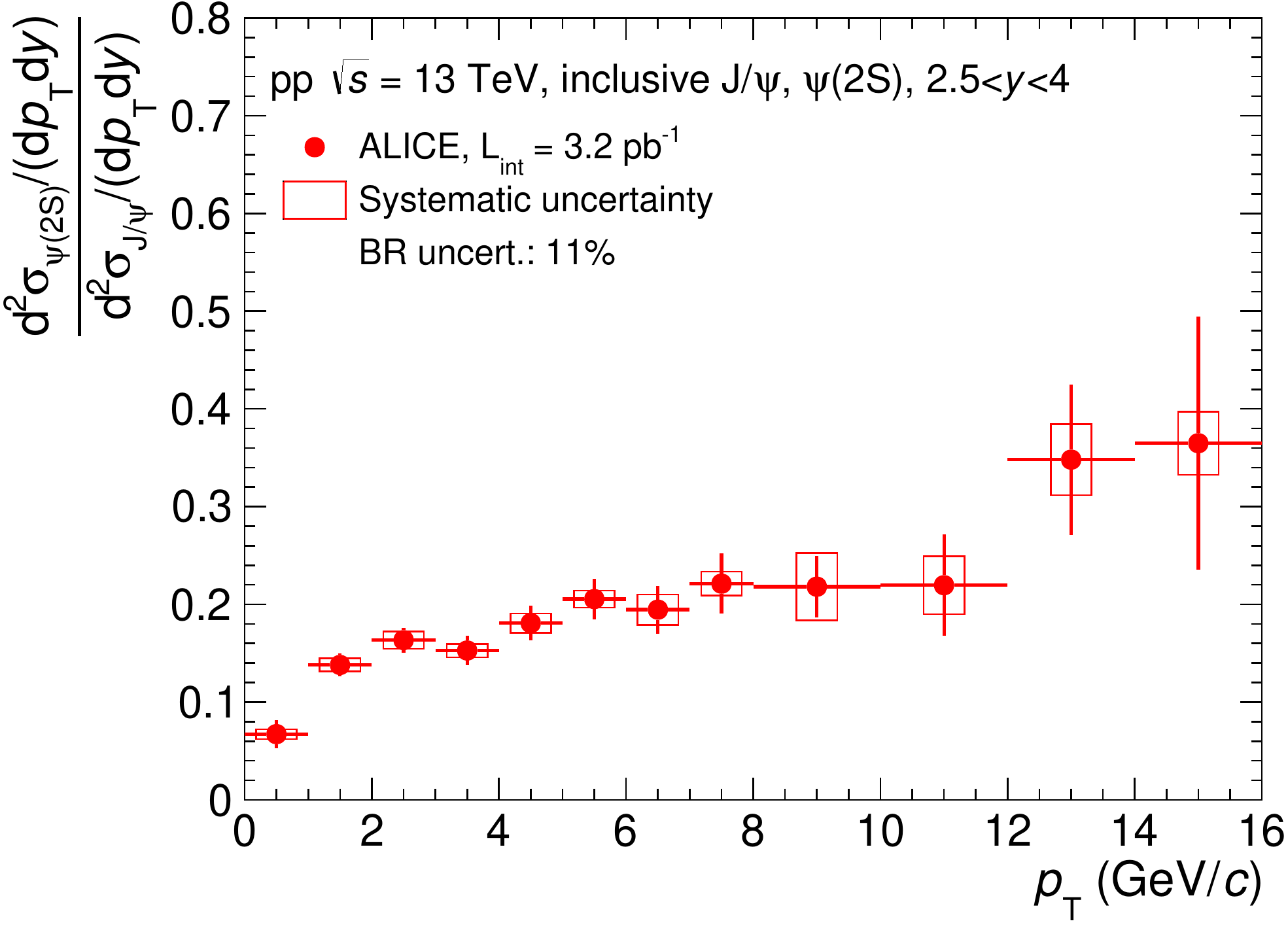}&\includegraphics[width=0.48\textwidth]{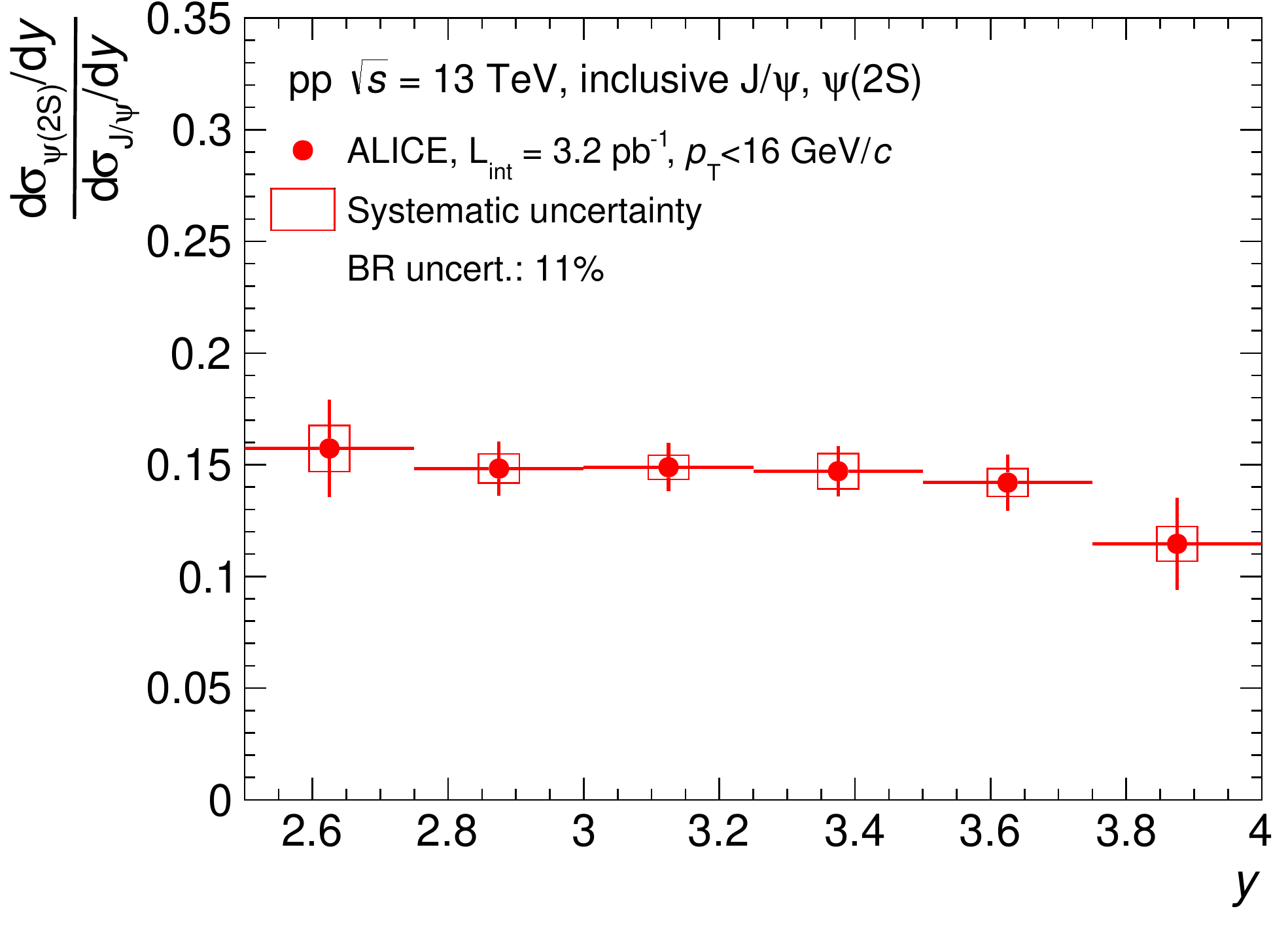}\\
\end{tabular}
\caption{\label{fig:InvYields13TeV}(color online). Inclusive $\jpsi$ cross sections (top), $\psip$ cross sections (middle) and $\psip$-to-$\jpsi$ cross section ratios (bottom) as a function of $\pt$~(left) and $\y$ (right) in pp collisions at $\sqrts=13$~TeV. $\jpsi$ cross sections are compared to LHCb measurements at the same $\sqrts$~\cite{Aaij:2015rla}. Open symbols are the reflection of the positive-$y$ measurements with respect to $y=0$.}
\end{figure}

%\subsection{$\jpsi$ and $\psip$ pp cross sections}
Fig.~\ref{fig:InvYields13TeV} summarizes the inclusive $\jpsi$ and $\psip$ cross sections measured by ALICE in pp collisions at $\sqrts=13$~TeV as a function of the charmonium $\pt$ (left column) and $y$ (right column). The top row shows the $\jpsi$ cross sections, middle row the $\psip$ cross sections and bottom row the $\psip$-to-$\jpsi$ cross section ratios. In all figures except Figs.~\ref{fig:meanpT} and~\ref{fig:Sigmaintegrated}, systematic uncertainties are represented by boxes, while vertical lines are used for statistical uncertainties. 

The $\jpsi$ production cross sections as a function of $\pt$ and $\y$ are compared to measurements published by LHCb~\cite{Aaij:2015rla} at the same energy. The quoted LHCb values correspond to the sum of the prompt and the non-prompt contributions to the $\jpsi$ production. For the comparison as a function of $\pt$, the provided double-differential ($\pt$ and $y$) cross sections are summed to match ALICE $y$ coverage. 
The measurements of the two experiments are consistent within $1\sigma$ of their uncertainties.
The ALICE measurement extends the $\pt$ reach from $14$~GeV/$c$ to $30$~GeV/$c$ with respect to the LHCb results. 
For the $\psip$ measurement, no comparisons are performed as this is the only measurement available to date at this energy and $y$ range. 

Systematic uncertainties on the signal extraction are reduced when forming the $\psip$-to-$\jpsi$ cross section ratios shown in the bottom panels of Fig.~\ref{fig:InvYields13TeV} due to correlations between the numerator and the denominator. All other sources of systematic uncertainties cancel except for the uncertainties on the MC input $\pt$ and $\y$ parametrizations. Measured ratios show a steady increase as a function of $\pt$ and little or no dependence on $y$ within uncertainties. This is also the case at lower $\sqrts$ as it will be discussed in the next section.

\begin{figure}[t!]
\centering
\begin{tabular}{cc}
\includegraphics[width=0.48\textwidth]{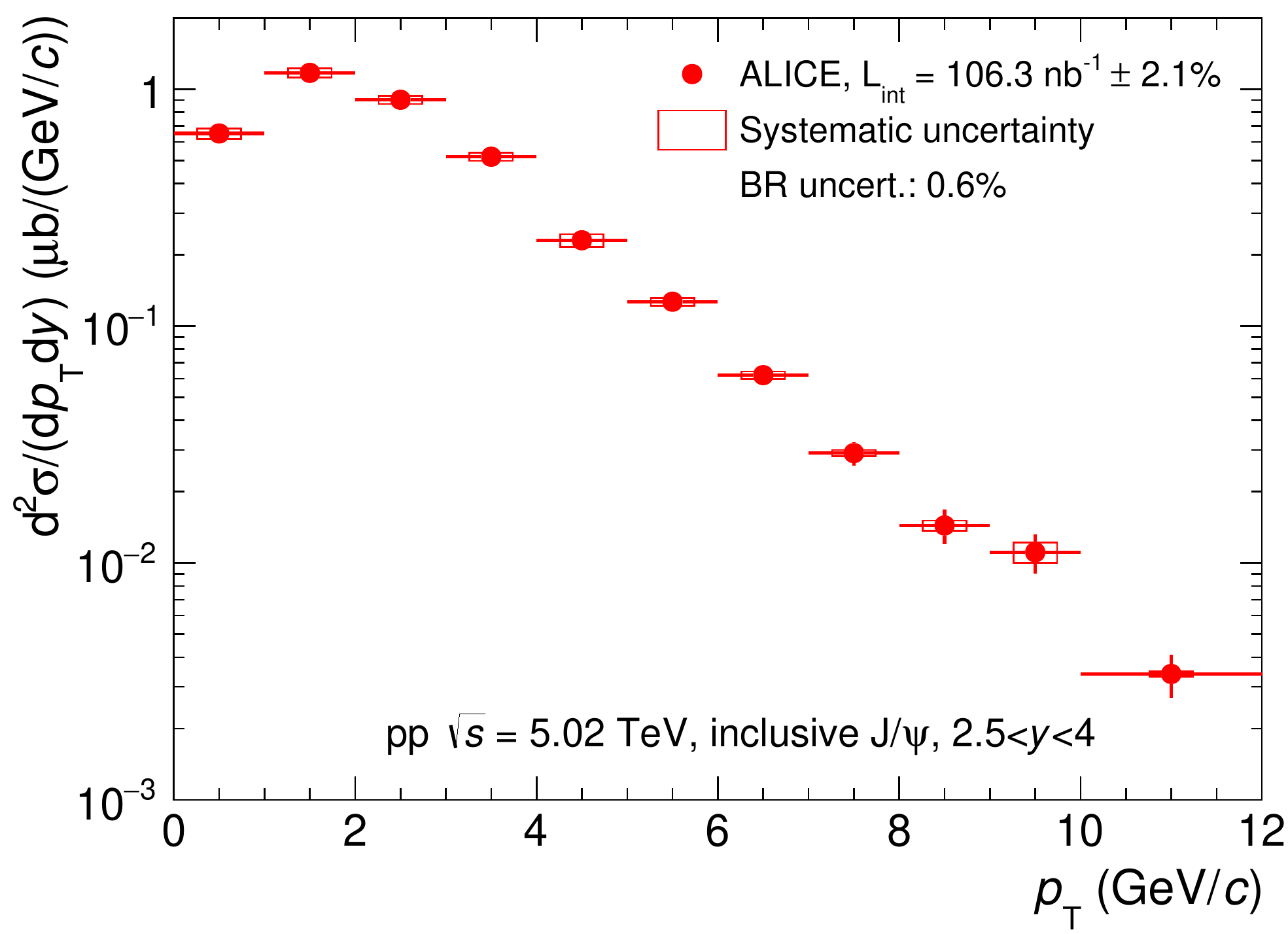}&\includegraphics[width=0.48\textwidth]{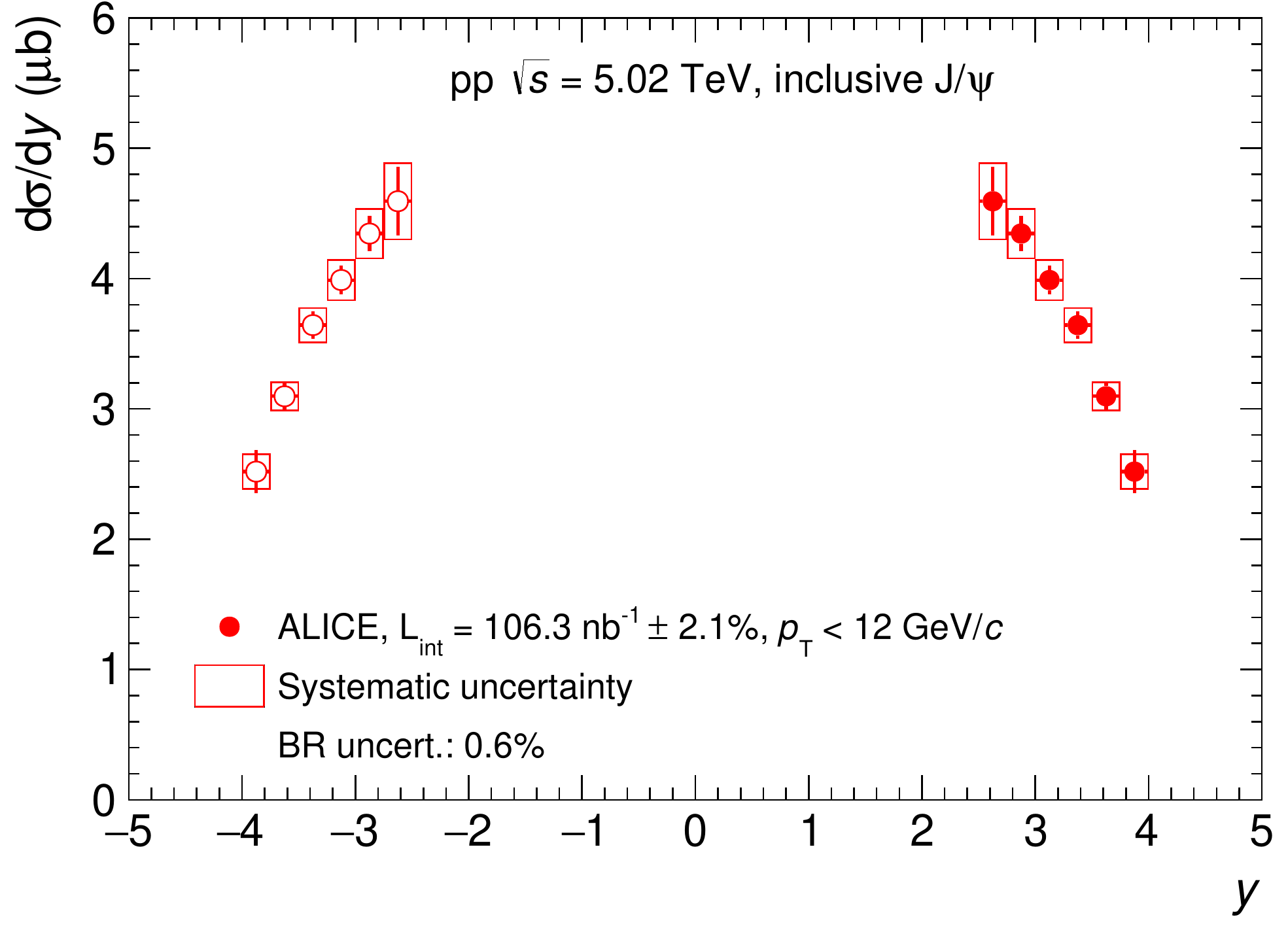}\\
\end{tabular}
\caption{\label{fig:InvYields5TeV}Inclusive $\jpsi$ cross sections as function of $\pt$~(left) and $\y$ (right) in pp collisions at $\sqrts=5.02$~TeV. Open symbols are the reflection of the positive-$y$ measurements with respect to $y=0$.}
\end{figure}

Fig.~\ref{fig:InvYields5TeV} shows the inclusive $\jpsi$ production cross section measurements performed by ALICE in pp collisions at $\sqrts=5.02$~TeV as a function of $\pt$ (left) and $y$ (right). The $\pt$-differential cross sections are published in~\cite{Adam:2016rdg} and serve as a reference for the $\jpsi$ nuclear modification factors in $\PbPb$ collisions at the same $\sqrts$. The $y$-differential cross sections are new to this analysis. Due to the limited integrated luminosity, only the $\pt$- and $y$-integrated $\psip$ cross section is measured using this data sample. It is discussed in the next section.

\subsection{Comparison to measurements at $\sqrts=2.76$, $7$ and $8$~TeV}

\begin{figure}[t!]
\centering
\begin{tabular}{cc}
\includegraphics[width=0.48\textwidth]{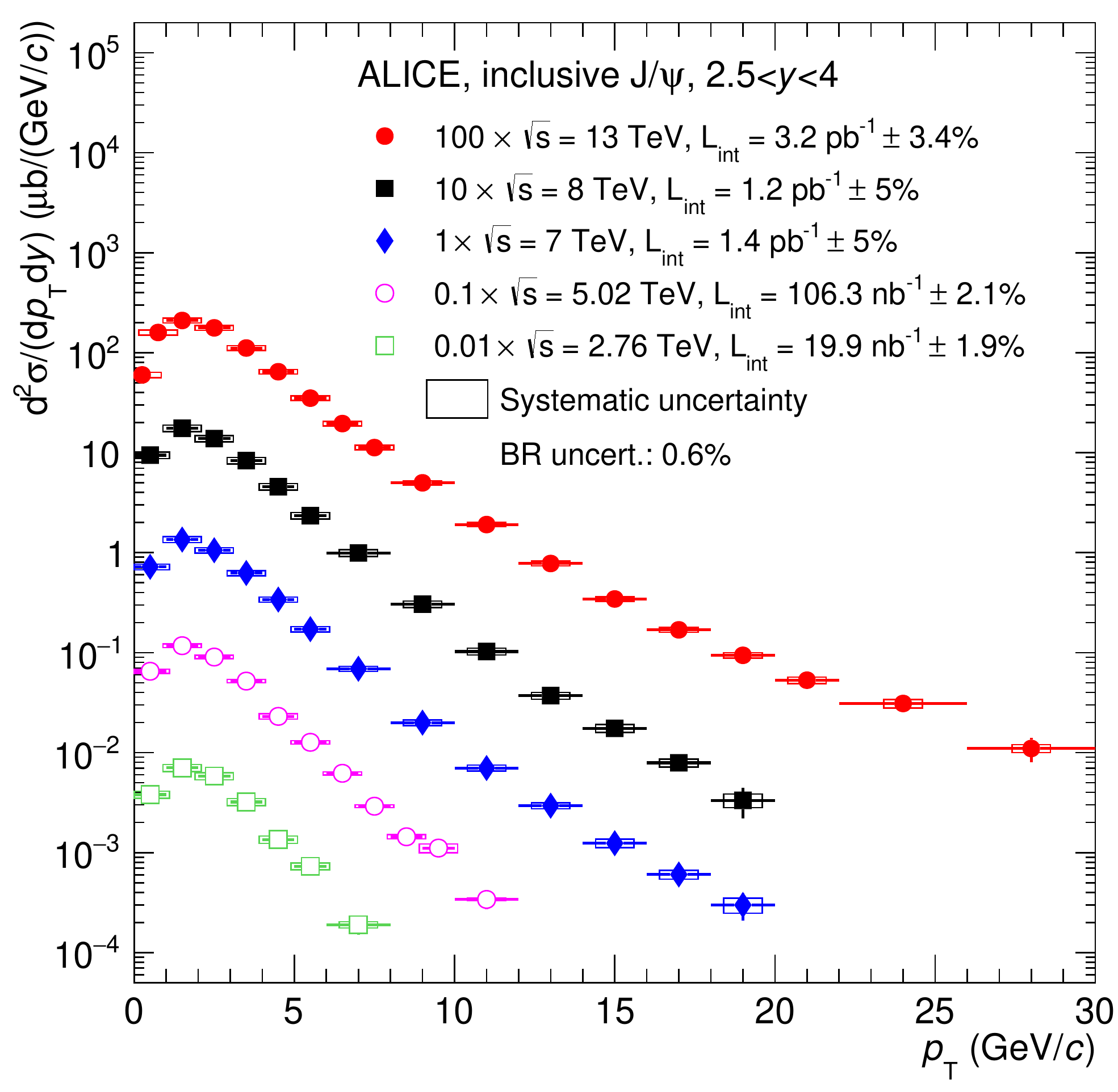}&\includegraphics[width=0.48\textwidth]{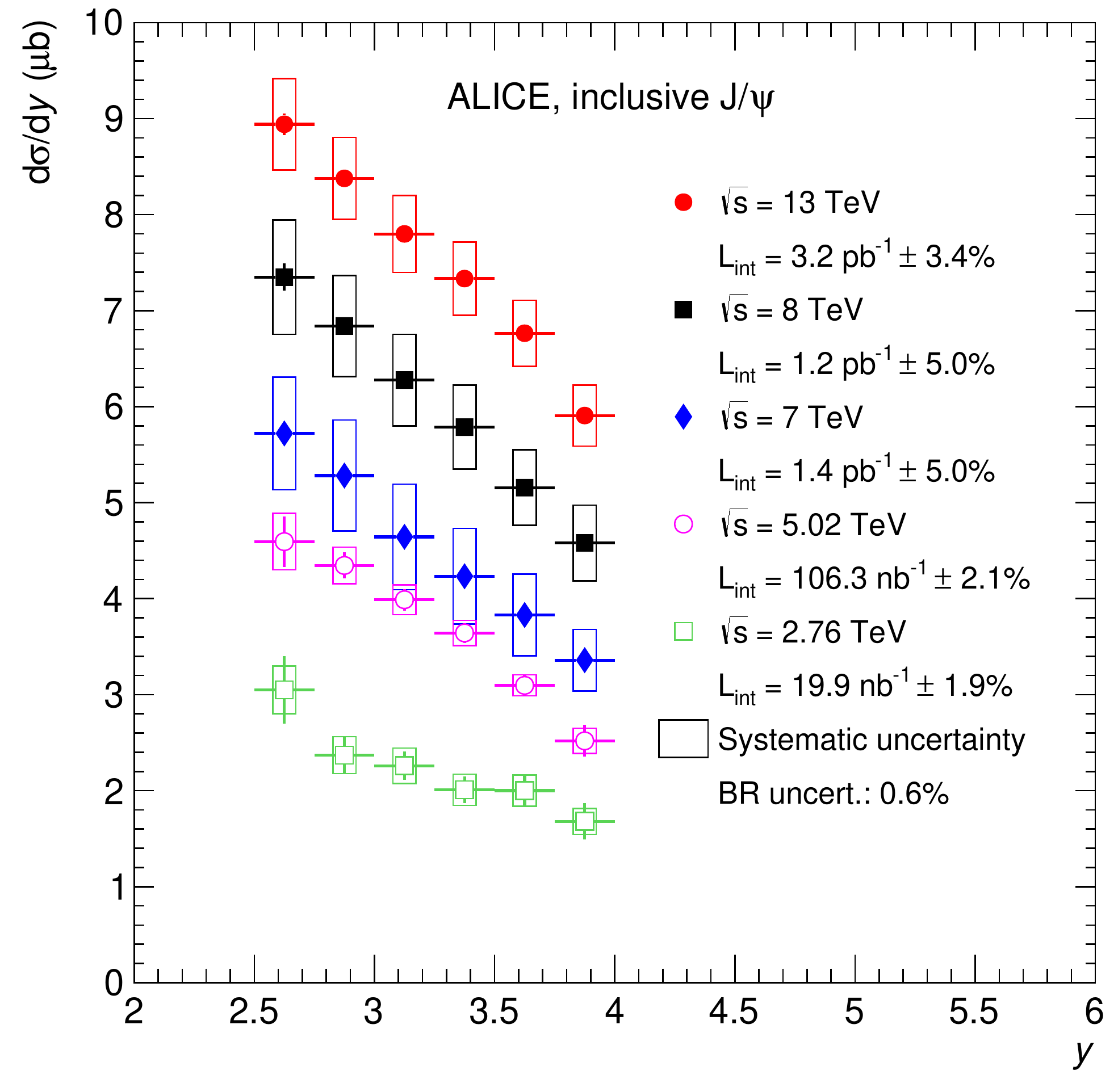}\\
\includegraphics[width=0.48\textwidth]{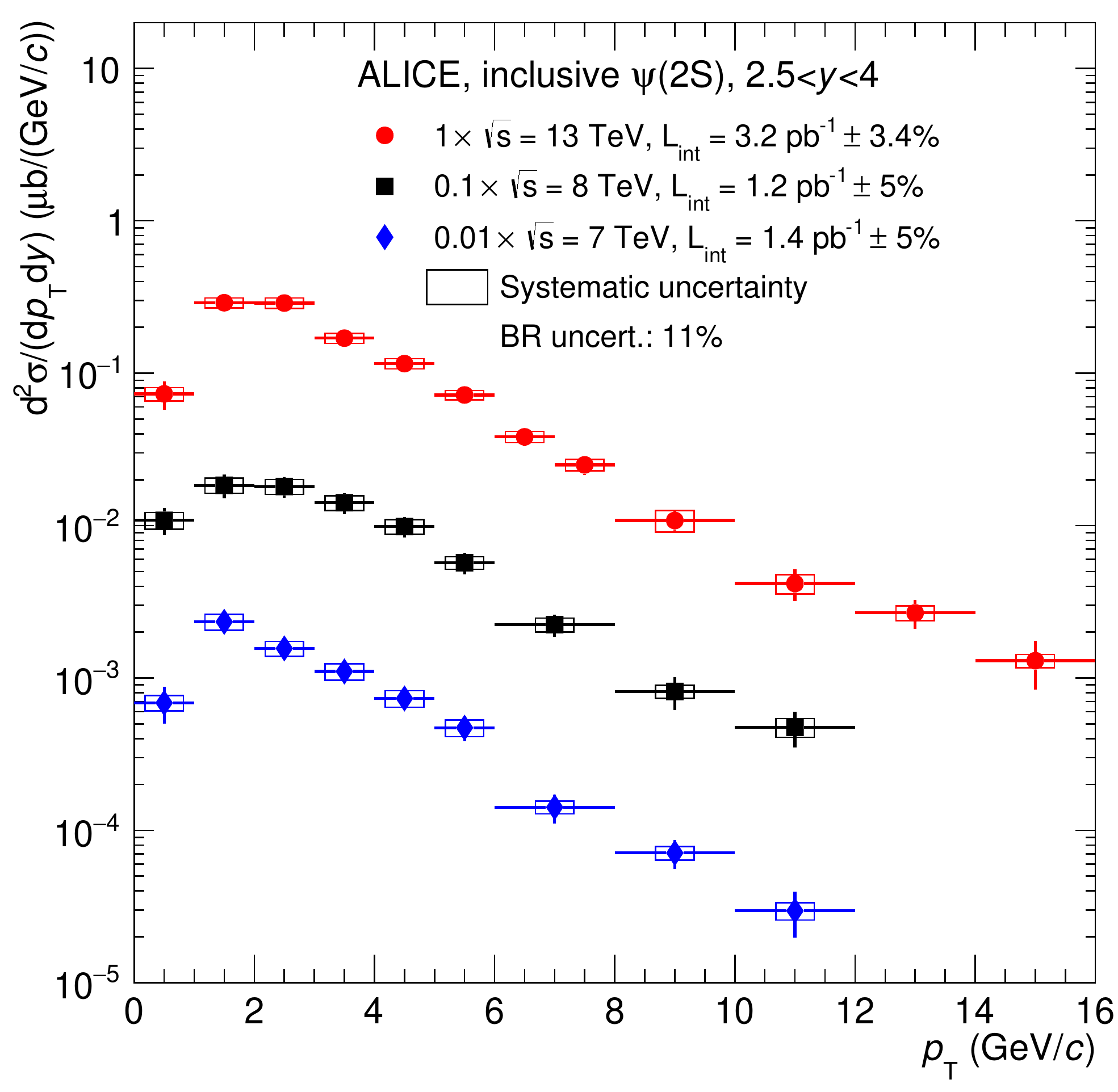}&\includegraphics[width=0.48\textwidth]{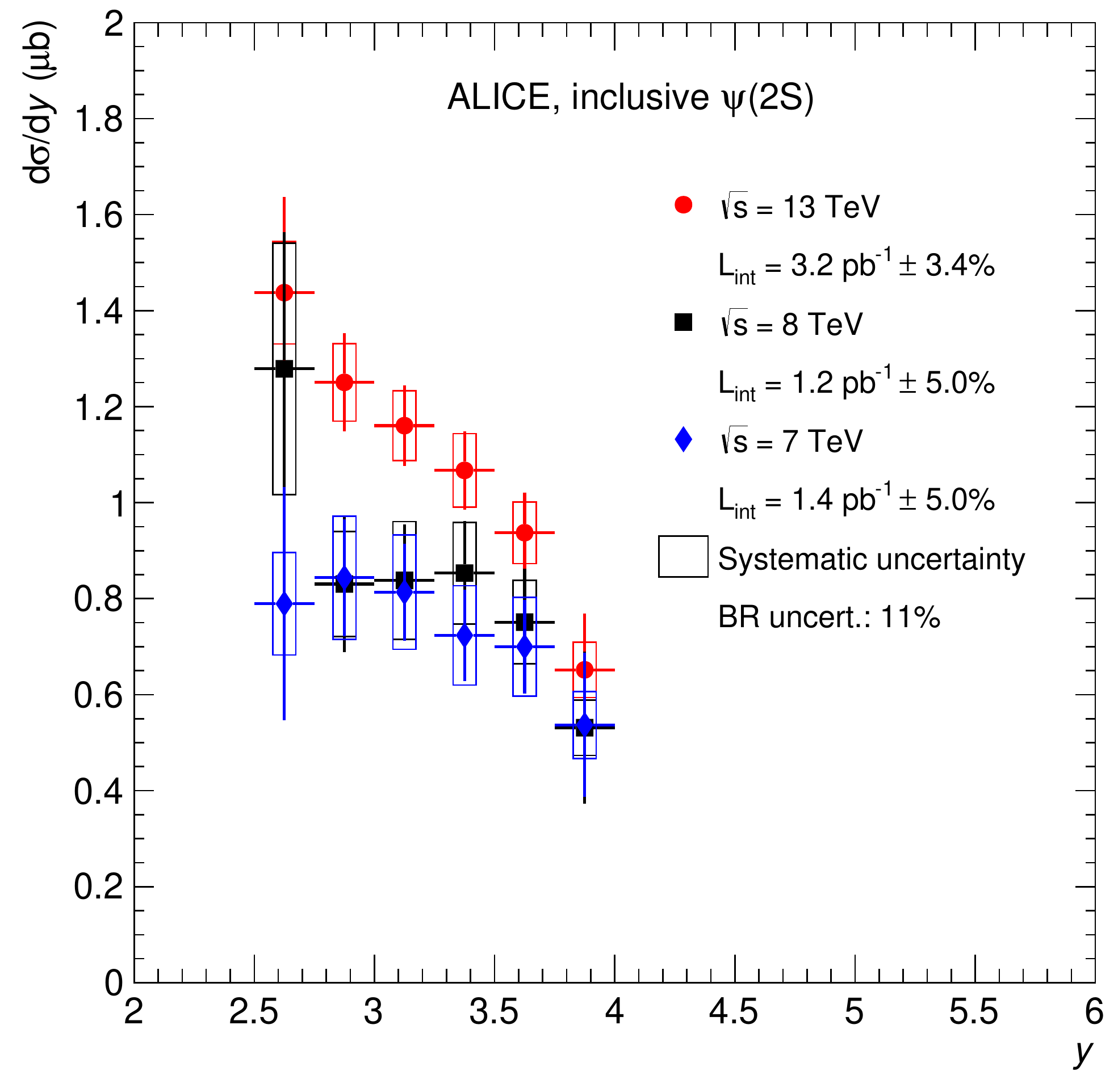}\\
\includegraphics[width=0.48\textwidth]{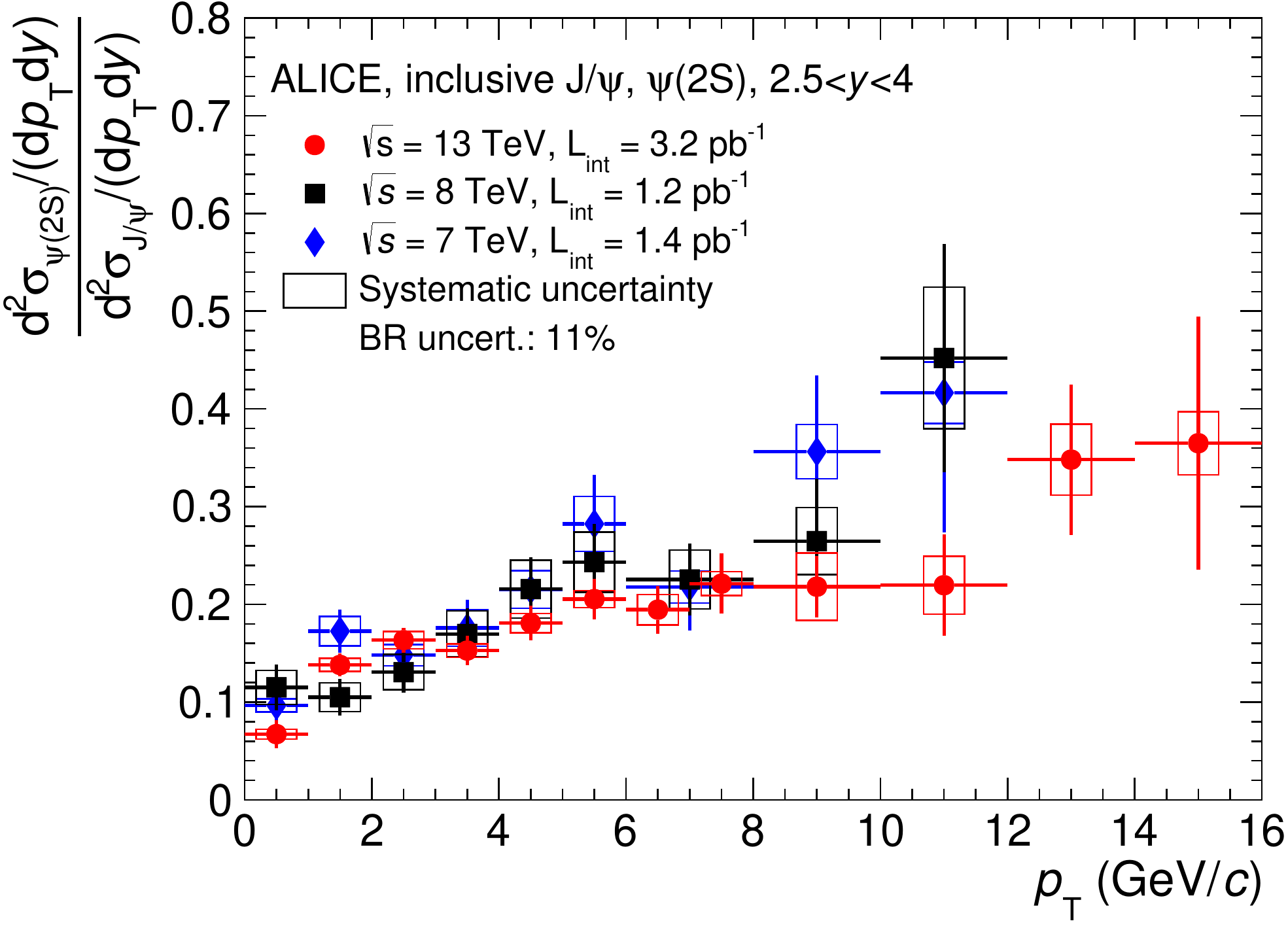}&\includegraphics[width=0.48\textwidth]{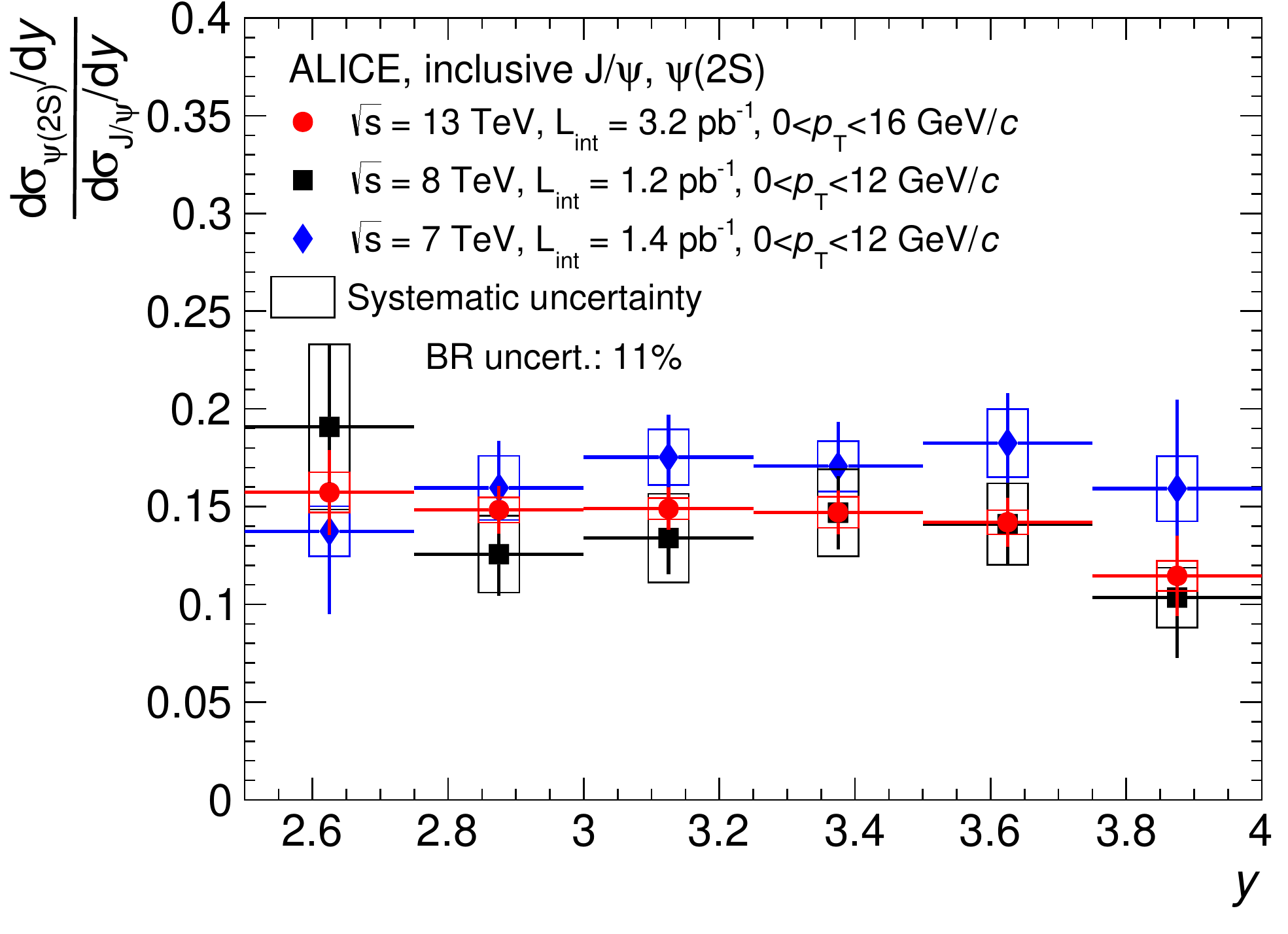}
\end{tabular}
\caption{\label{fig:InvYieldsAll}(color online). Inclusive $\jpsi$ cross sections (top), $\psip$ cross sections (middle) and $\psip$-to-$\jpsi$ cross section ratios (bottom) as function of $\pt$~(left) and $\y$ (right) in pp collisions at several values of $\sqrts$.}
\end{figure}

In Fig.~\ref{fig:InvYieldsAll}, the cross sections and cross section ratios presented in the previous section are compared to other forward-$y$ measurements in pp collisions at $\sqrts=2.76$~\cite{Abelev:2012kr}, 7~\cite{Abelev:2014qha} and 8~TeV~\cite{Adam:2015rta}. 
We note that the integrated luminosity used for each measurement increases almost systematically with increasing $\sqrts$, starting from $19.9$~nb$^{-1}$ at $\sqrts=2.76$~TeV up to $3.2$~pb$^{-1}$ at $\sqrts=13$~TeV. This, combined with the fact that the charmonium cross-section also increases with $\sqrt{s}$, has allowed to reach increasingly higher values of $\pt$ for both $\jpsi$ and $\psip$ measurements. For the $\jpsi$ this corresponds to an increase of the $\pt$ reach from $8$~GeV/$c$ at $\sqrts=2.76$~TeV up to $30$~GeV/$c$ at $\sqrts=13$~TeV. For the $\psip$ the corresponding increase goes from $12$~GeV/$c$ at $\sqrts=7$~TeV to $16$~GeV/$c$ at $\sqrts=13$~TeV. 

The $\jpsi$ $\pt$-differential cross section measurements shown in the top-left panel of Fig.~\ref{fig:InvYieldsAll} indicate a hardening of the spectra with increasing $\sqrts$. Also, for $\sqrts\geq7$~TeV, a change in the slope of the $\pt$-differential cross section is visible for $\pt > 10$~GeV/$c$. This change in slope is attributed to the onset of the contribution from non-prompt $\jpsi$ to the inclusive cross section as it will be discussed in Sec.~\ref{sec:models}. 

The corresponding $\psip$ differential cross section measurements are shown in the middle panels of Fig.~\ref{fig:InvYieldsAll}. The smaller cross sections with respect to $\jpsi$ result in a smaller $\pt$ reach as well as larger statistical uncertainties as a function of both $\pt$ (left panel) and $y$ (right panel).

In the bottom panels of Fig.~\ref{fig:InvYieldsAll} the measured $\psip$-to-$\jpsi$ cross section ratios are compared as a function of $\pt$ (left) and $y$ (right) for pp collisions at $\sqrt{s}=7$, $8$ and $13$~TeV. No significant change neither in shape nor magnitude of the ratio is observed among the three energies within the current uncertainties.

\begin{figure}[t!]
\centering
\begin{tabular}{cc}
\includegraphics[width=0.48\textwidth]{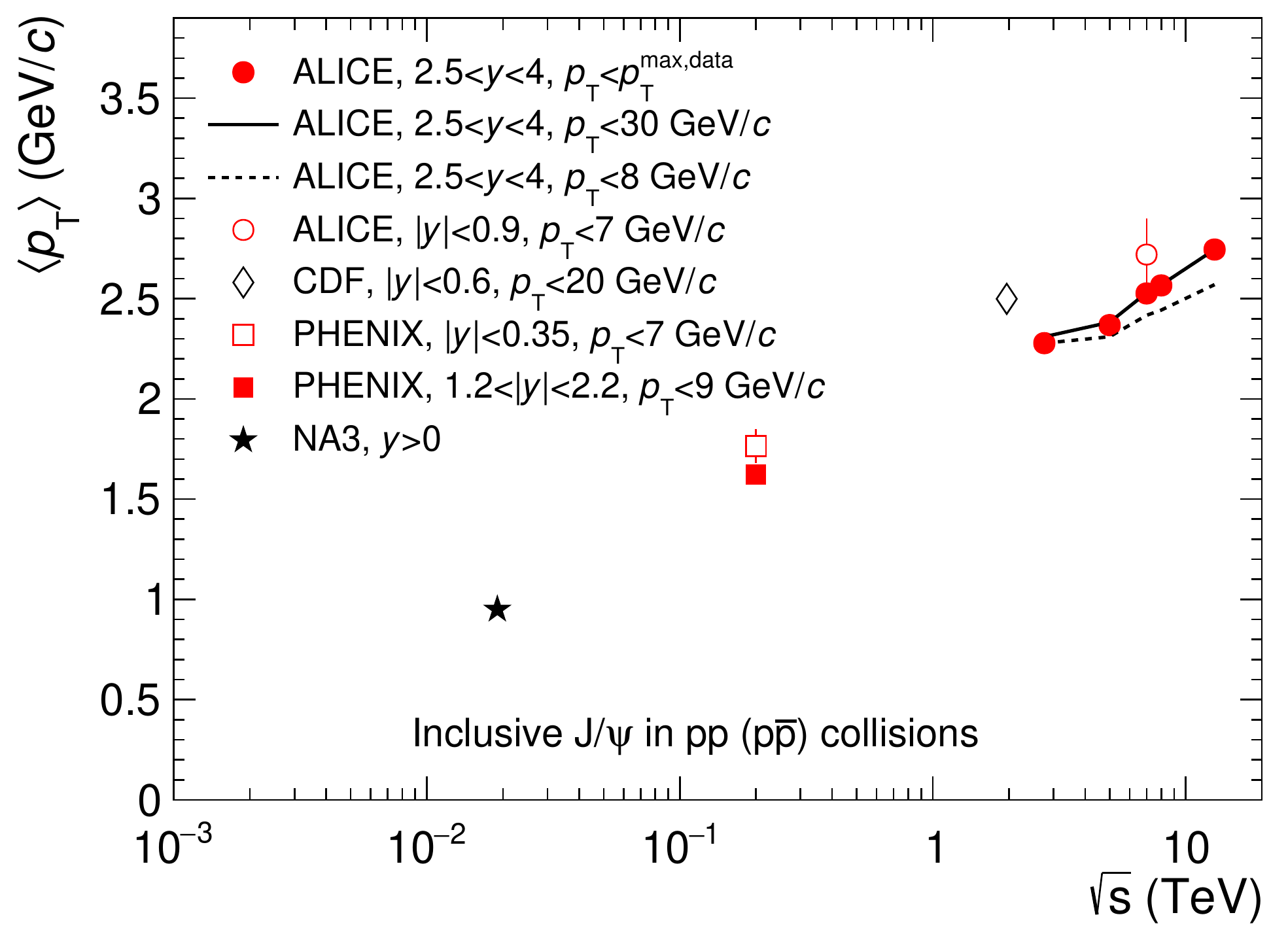}&\includegraphics[width=0.48\textwidth]{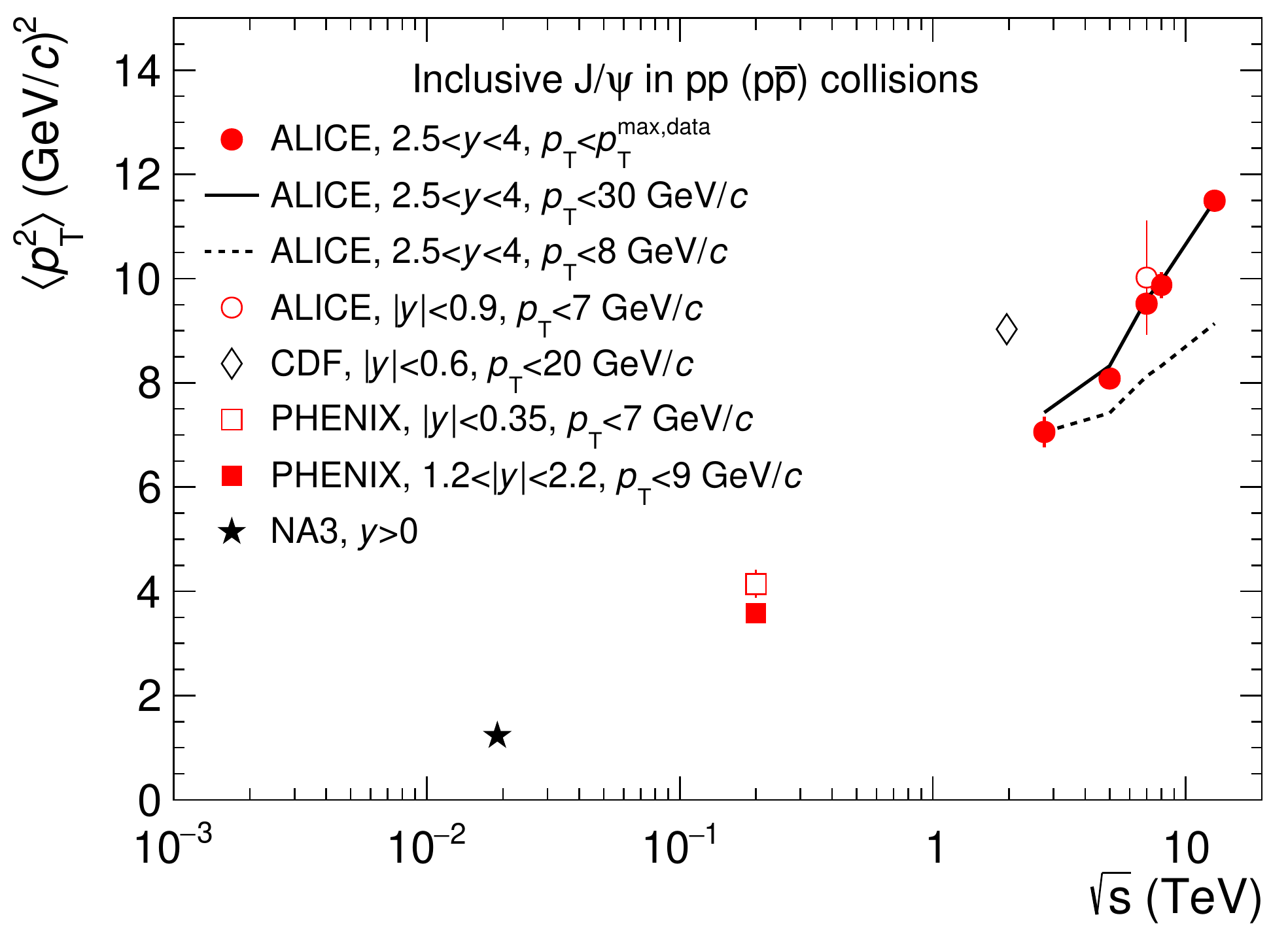}\\
\includegraphics[width=0.48\textwidth]{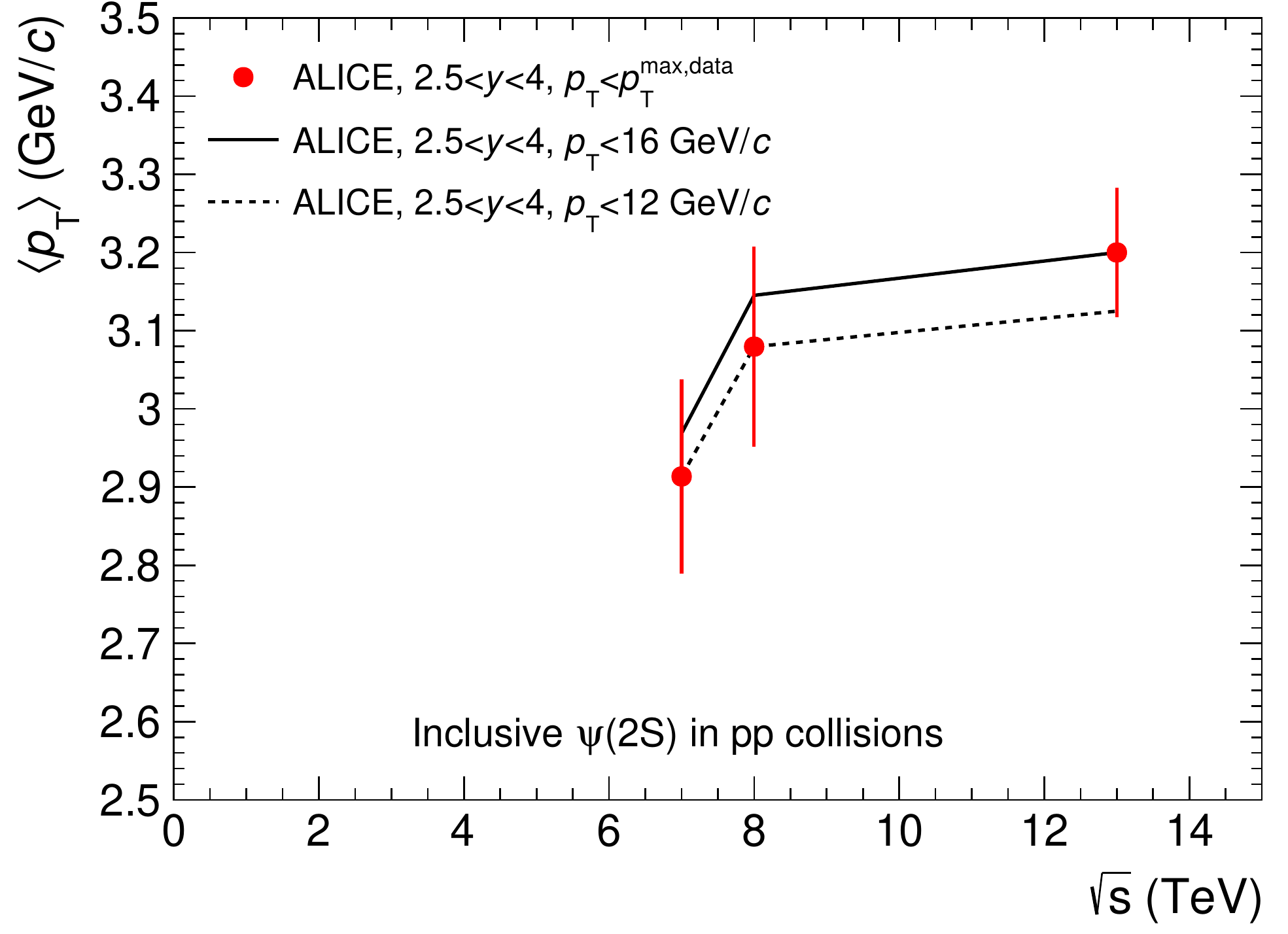}&\includegraphics[width=0.48\textwidth]{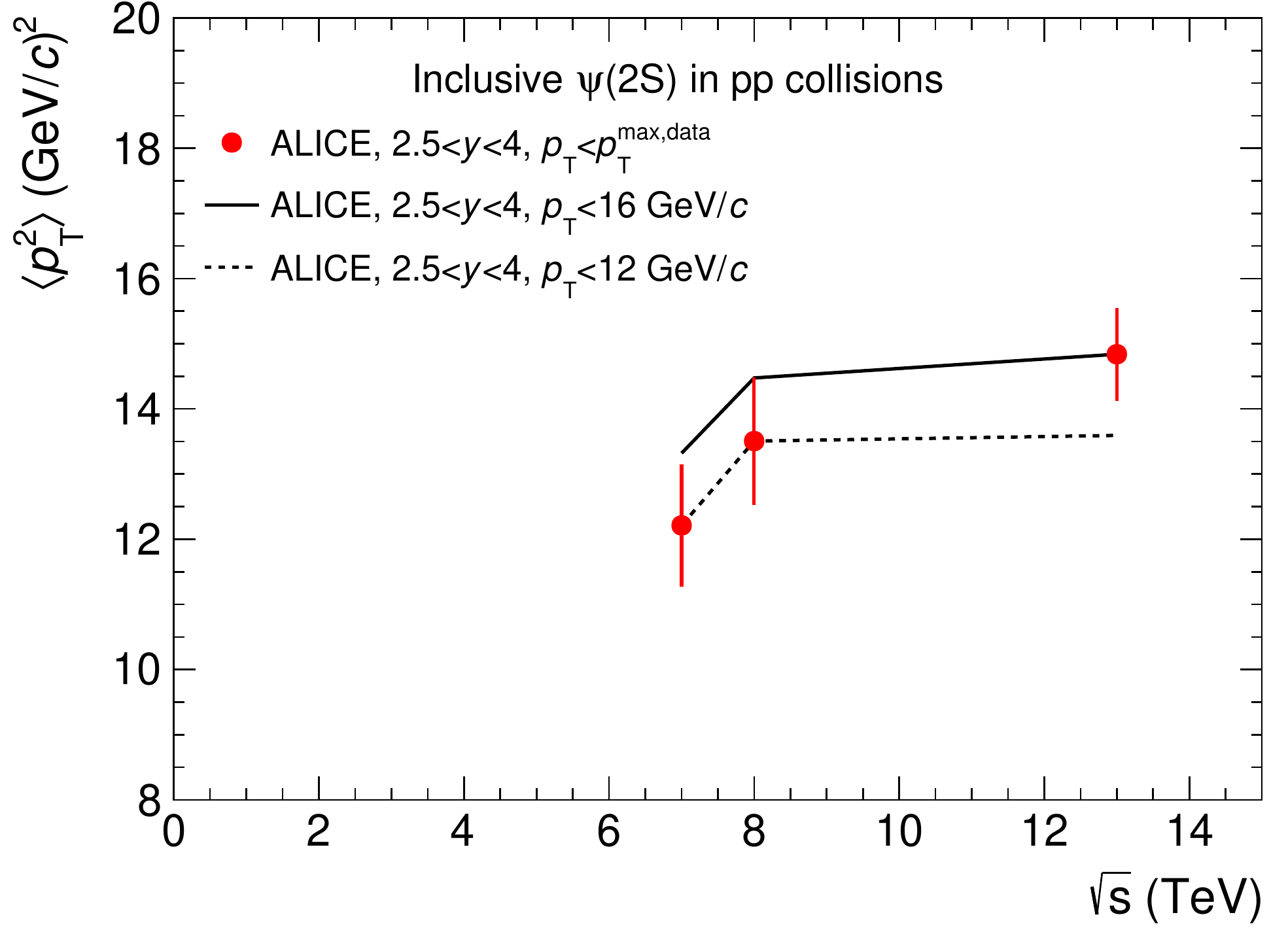}
\end{tabular}
\caption{\label{fig:meanpT}$\meanpt$ (left) and $\meanptsq$ (right) as a function of $\sqrts$ for $\jpsi$~(top) and $\psip$ (bottom). Circles correspond to ALICE data, while the other symbols correspond to measurements at lower $\sqrts$. 
Vertical lines around the data points correspond to the quadratic sum of the statistical and uncorrelated systematic uncertainties.
The solid lines correspond to calculating $\meanpt$ and $\meanptsq$ when extrapolating the $\pt$ coverage to the largest available range in ALICE data ($0<\pt<30$~GeV/$c$ for $\jpsi$ and $0<\pt<16$~GeV/$c$ for $\psip$), while the dashed lines correspond to truncating the data to the smallest $\pt$ range available ($0<\pt<8$~GeV/$c$ for $\jpsi$ and $0<\pt<12$~GeV/$c$ for $\psip$).}
\end{figure}

To better quantify the hardening of the $\jpsi$ and $\psip$ $\pt$ spectra with increasing $\sqrts$, a computation of the corresponding mean transverse momentum $\meanpt$ and mean transverse momentum square $\meanptsq$ is performed. This is achieved by fitting the $\jpsi$ and $\psip$ $\pt$-differential cross sections with the following function:

\begin{equation}
f(\pt) = C \times \frac{ \pt }{ {\left( 1 + {\left( \frac{\pt}{ p_0 } \right)}^2 \right)}^n },
\label{eq_meanpt}
\end{equation}
with the parameters $C$, $p_0$ and $n$ left free.

The $\meanpt$ and $\meanptsq$ are then obtained as the first and second moments of the above function in a given $\pt$ range. The uncertainty on these quantities is evaluated by multiplying the covariance matrix of the fit on each side by the relevant Jacobian matrix, evaluated numerically and taking the square root of the result. This is performed either considering separately the statistical and uncorrelated systematic uncertainties, or by using their quadratic sum in order to obtain the corresponding statistical, systematic or total uncertainty. A similar approach was adopted in~\cite{Abelev:2012kr}.

\begin{figure}[t!]
\centering
\begin{tabular}{cc}
\includegraphics[width=0.48\textwidth]{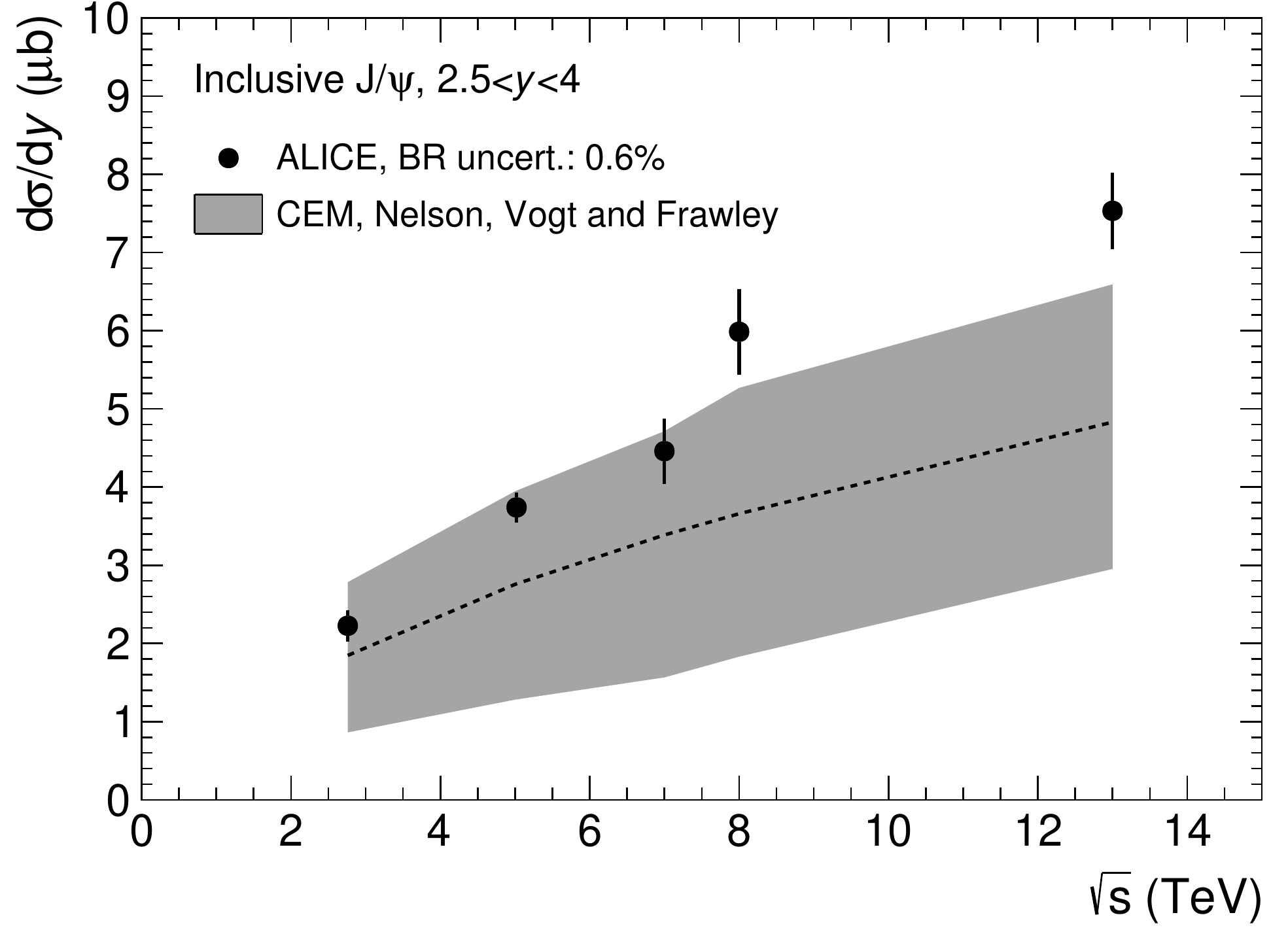}&
\includegraphics[width=0.48\textwidth]{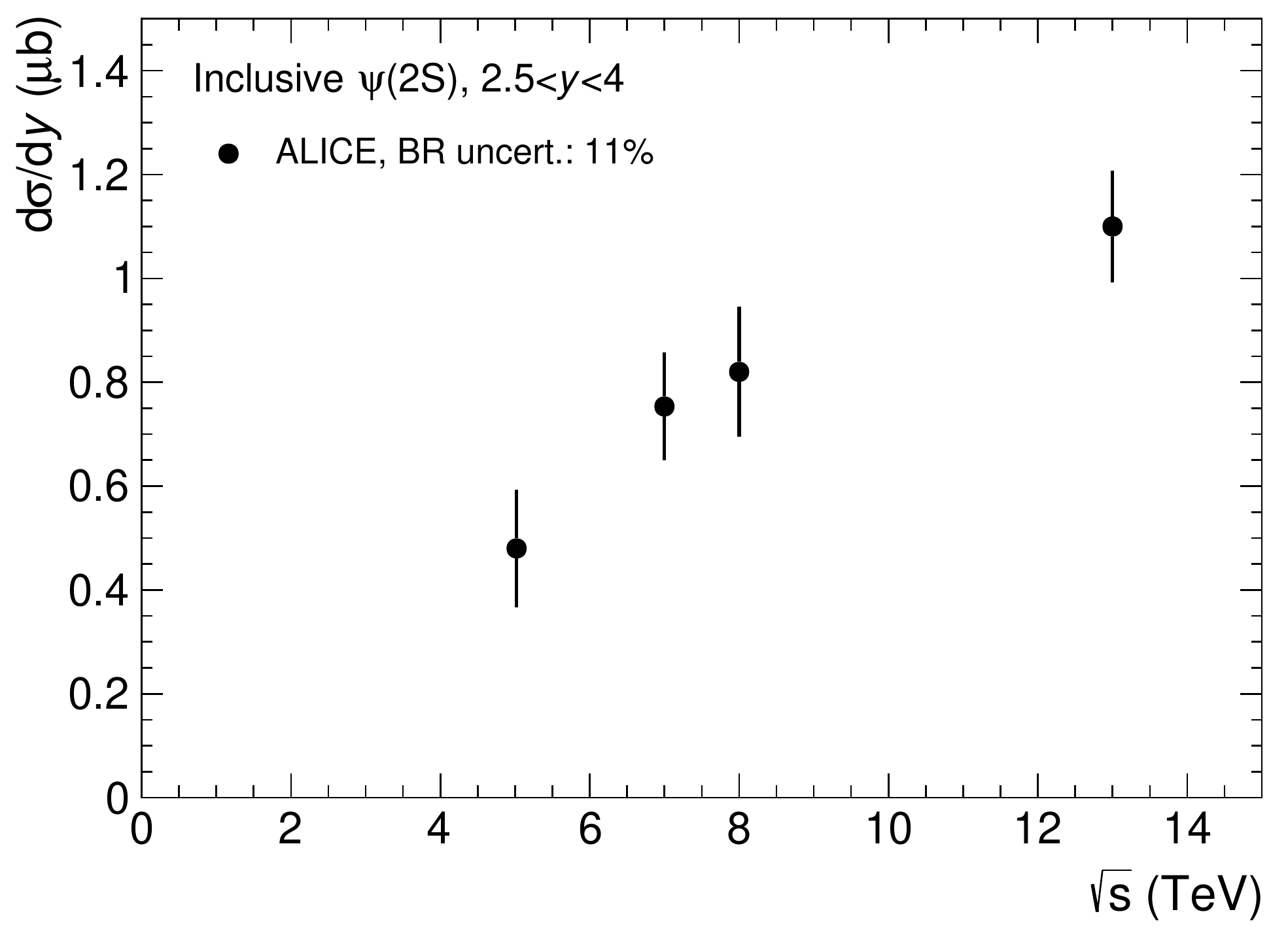}
\end{tabular}
\caption{\label{fig:Sigmaintegrated}$\jpsi$ (left) and $\psip$ (right) inclusive cross section $\dif\sigma/\dif y$ as a function of $\sqrts$. Vertical lines correspond to the quadratic sum of the statistical and uncorrelated systematic uncertainties. $\jpsi$ cross sections are compared to a CEM calculation from~\cite{Nelson:2012bc}. }
\end{figure}

Fig.~\ref{fig:meanpT} shows the $\meanpt$ (left) and $\meanptsq$~(right) results for $\jpsi$ (top) and $\psip$ (bottom). In this figure as well as in Fig.~\ref{fig:Sigmaintegrated}, the vertical lines correspond to the quadratic sum of the statistical and uncorrelated systematic uncertainties.

For $\jpsi$ at $\sqrts=2.76$~TeV the value from~\cite{Abelev:2012kr} is used. At $\sqrts=7$~TeV the data from~\cite{Abelev:2014qha} are used instead of the result from~\cite{Abelev:2012kr} because the available integrated luminosity is much larger ($\times 90$) and the $\pt$ reach increased from $8$ to $20$~GeV/$c$. It was checked that both results are consistent when truncated to the same $\pt$ range. At $\sqrts=8$~TeV the data from~\cite{Adam:2015rta} are used, while for $\sqrts=5.02$ and $13$~TeV the results are from this analysis.

In the top panels of Fig.~\ref{fig:meanpT}, ALICE measurements are compared to lower energy results from CDF~\cite{Acosta:2004yw}, PHENIX~\cite{Adare:2006kf} and NA3~\cite{Badier:1983dg}. A steady increase of $\meanpt$ and $\meanptsq$ is observed with increasing $\sqrts$. This is consistent with the expected hardening of the corresponding $\pt$ distributions. Moreover, values at mid- are systematically larger than at forward-rapidity. As discussed in~\cite{Adare:2006kf}, this observation could be attributed to an increase in the longitudinal momentum at forward-rapidity leaving less energy available in the transverse plane.
The bottom panels of Fig.~\ref{fig:meanpT} show the corresponding measurements for $\psip$ at $\sqrts=7$, $8$ and $13$~TeV. An increase with $\sqrts$ is also observed similar to that of the $\jpsi$. 

Part of the increase observed for ALICE measurements shown in all four panels of Fig.~\ref{fig:meanpT} is due to the fact that the $\pt$ range used for evaluating $\meanpt$ and $\meanptsq$, chosen to be the same as in the corresponding data, also increases with $\sqrts$. 
To illustrate this effect, these quantities were re-calculated either when truncating the data to the smallest available $\pt$ range ($0<\pt<8$~GeV/$c$ for $\jpsi$ and $0<\pt<12$~GeV/$c$ for $\psip$) or when using the fit based on Eq.~\ref{eq_meanpt} to extrapolate the data to the largest available range ($0<\pt<30$~GeV/$c$ for $\jpsi$ and $0<\pt<16$~GeV/c for $\psip$). The resulting values are shown in the figures as dashed lines for the truncation and solid lines for the extrapolation. In all cases the observed increasing trend still holds.

Finally, Fig.~\ref{fig:Sigmaintegrated} shows the $\jpsi$ (left) and $\psip$ (right) $\pt$- and $y$-integrated inclusive cross sections as a function of $\sqrts$, measured by ALICE in the $y$ range $2.5<y<4$. For both particles a steady increase of ${\rm d}\sigma/{\rm d}y$ is observed as a function of increasing $\sqrts$. For the $\jpsi$, the cross sections are compared to a calculation done by Nelson, Vogt and Frawley in the CEM framework~\cite{Nelson:2012bc}. While the data and the model are compatible within uncertainties, the data lie on the upper side of the calculation and the difference to the central value becomes larger with increasing $\sqrts$.

%%%%%%%%%%%%%%%%%%%%%%%%%%%%%%%%%%%%%%%%%%%%%%%
\subsection{\label{sec:models}Comparisons to models}

\begin{figure}[t!]
\centering
\begin{tabular}{cc}
\includegraphics[width=0.48\textwidth]{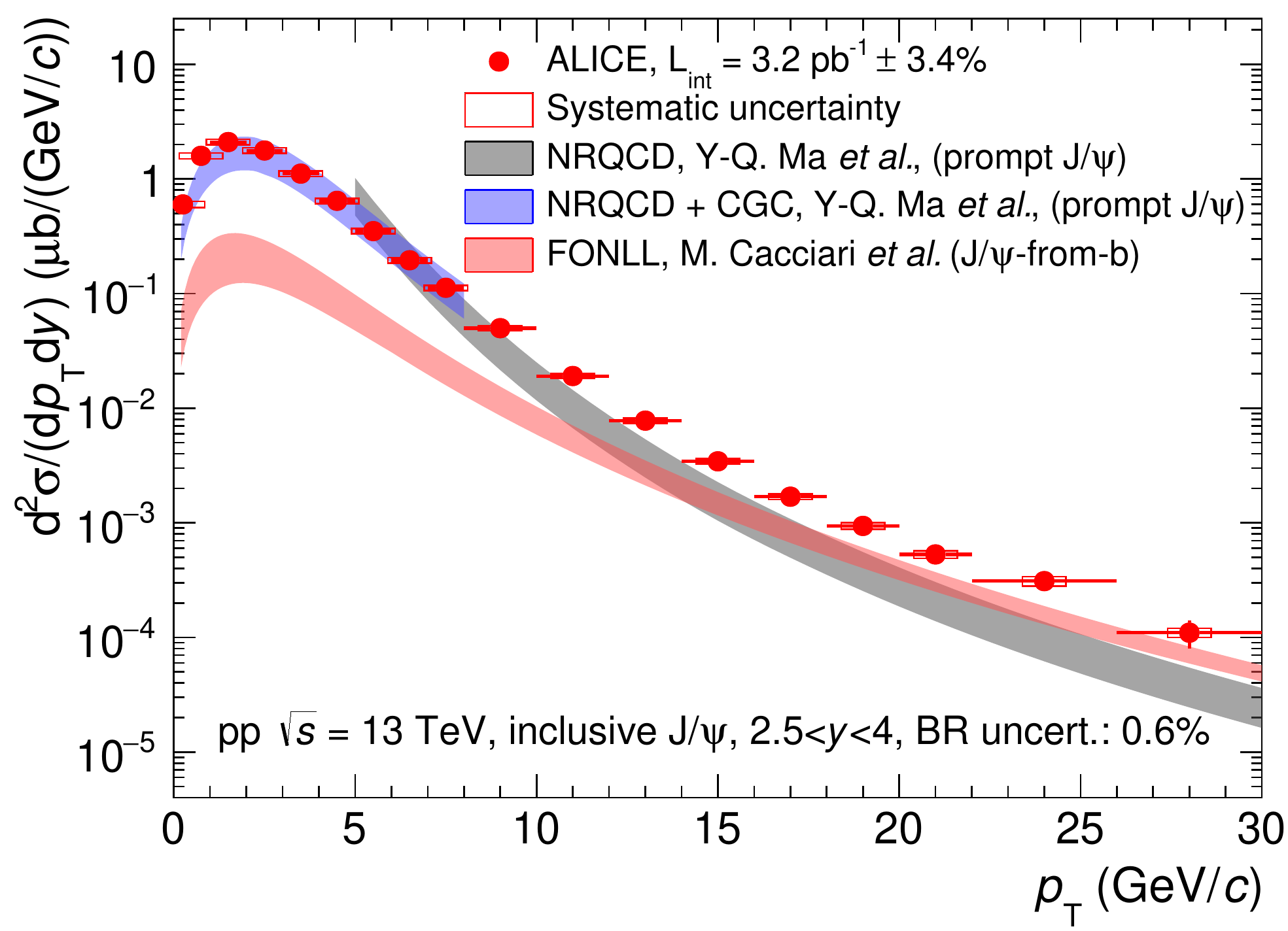}&\includegraphics[width=0.48\textwidth]{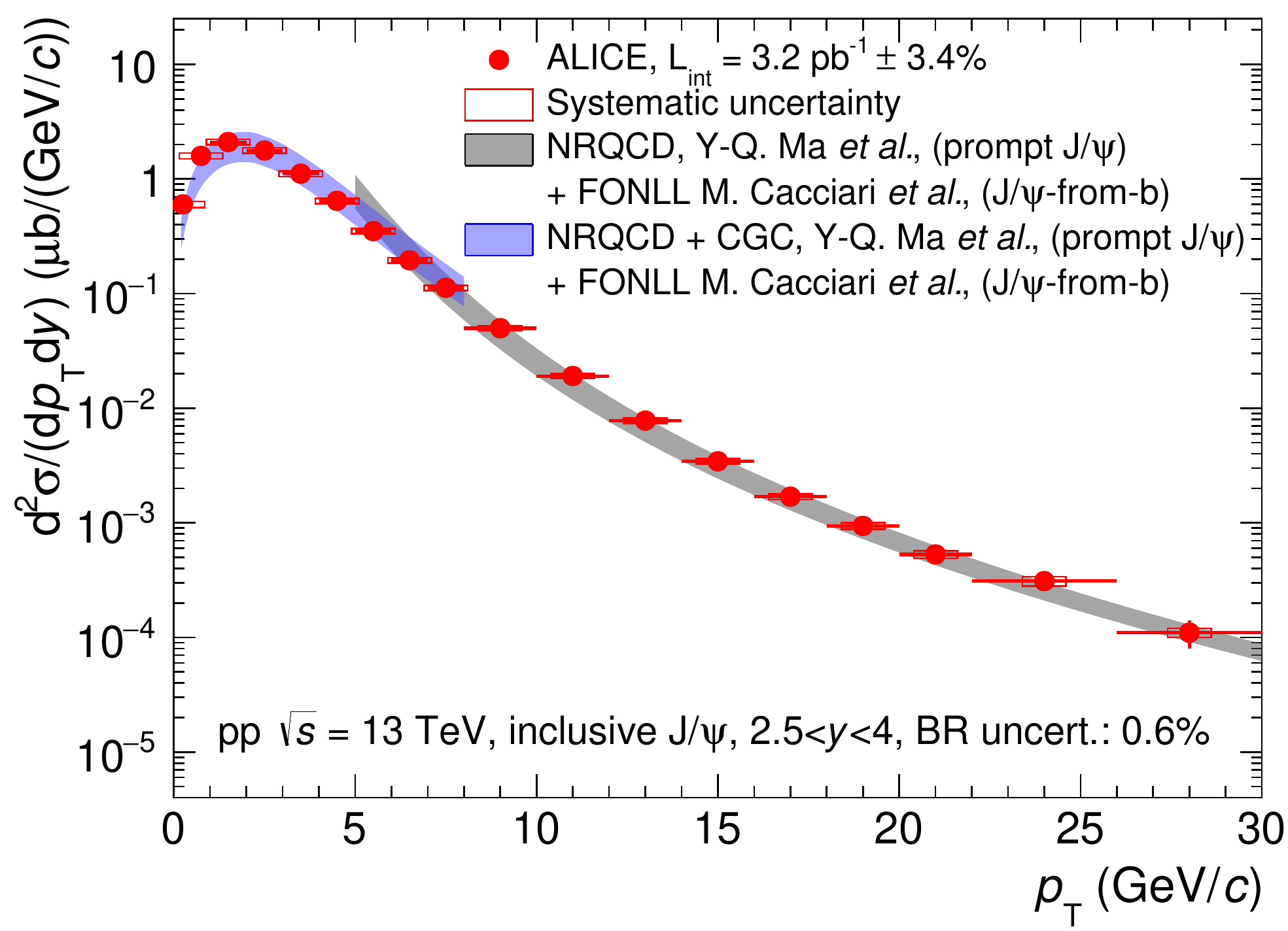}\\
\end{tabular}
\caption{\label{fig:PtMa13TeV}(color online). Left panel: $\jpsi$ differential cross sections (red circles) in pp collisions at $\sqrts=13$~TeV compared to NLO NRQCD (grey)~\cite{Ma:2010yw}, LO NRQCD coupled with CGC (blue)~\cite{Ma:2014mri} and FONLL (red)~\cite{Cacciari:2012ny}. Right panel: the non-prompt $\jpsi$ contribution estimated with FONLL is summed to the two calculations for prompt $\jpsi$ production and compared to the same data.}
\end{figure}

As discussed in the introduction, all ALICE $\jpsi$ and $\psip$ measurements presented in this paper are inclusive and consist of a prompt and a non-prompt contribution. 
In order to compare model calculations to the data both contributions must be accounted for. This is illustrated in Fig.~\ref{fig:PtMa13TeV} for the $\jpsi$ production cross section as a function of $\pt$ in pp collisions at $\sqrts=13$~TeV. 

In the left panel of Fig.~\ref{fig:PtMa13TeV}, ALICE data are compared to three calculations: 
\begin{enumerate*}[label=(\roman*)]
\item in grey to a prompt $\jpsi$ Next-to-Leading-Order (NLO) NRQCD calculation from Ma, Wang and Chao~\cite{Ma:2010yw}, 
\item in blue to a prompt $\jpsi$ Leading Order (LO) NRQCD calculation coupled to a Color Glass Condensate (CGC) description of the low-$x$ gluons in the proton from Ma and Venugopalan~\cite{Ma:2014mri} and 
\item in red to a non-prompt $\jpsi$ Fixed-Order Next-to-Leading Logarithm (FONLL) calculation by Cacciari {\em et al.}~\cite{Cacciari:2012ny}.
\end{enumerate*}
Both NRQCD prompt $\jpsi$ calculations account for the decay of $\psip$ and $\chi_c$ into $\jpsi$.

For $\pt<8$~GeV/$c$ where the contribution from non-prompt $\jpsi$ estimated using FONLL is below 10\%, the NRQCD+CGC prompt $\jpsi$ calculation reproduces the data reasonably well. For higher $\pt$ on the other hand, the NLO NRQCD calculation underestimates the measured cross sections and the disagreement increases with increasing $\pt$. This disagreement is explained by the corresponding increase of the non-prompt $\jpsi$ contribution, which according to FONLL, becomes as high as the prompt contribution and even exceeds it for $\pt>15$~GeV/$c$. This is consistent with the measured non-prompt $\jpsi$ fractions reported by LHCb in~\cite{Aaij:2015rla}.

In the right panel of Fig.~\ref{fig:PtMa13TeV}, the NRQCD and FONLL calculations for prompt and non-prompt $\jpsi$ production are summed in order to obtain an {\em ad hoc} model of inclusive $\jpsi$ production. The sum is performed separately for the NRQCD+CGC calculation at low $\pt$ and the NLO NRQCD at high $\pt$. 
In both cases, the uncertainties on FONLL and NRQCD are considered as uncorrelated when calculating the uncertainty band on the sum.
This is motivated by the fact that the NRQCD calculations refer to the production of charm quarks and charmed mesons, while the FONLL calculation applies to the production of beauty quarks and $b$-hadrons which are then decayed into $\jpsi$ mesons. A good description of the data is obtained over the full $\pt$ range and spanning more than four orders of magnitude in the cross sections. 

\begin{figure}[t!]
\renewcommand{\arraystretch}{0.5}
\includegraphics[width=\textwidth]{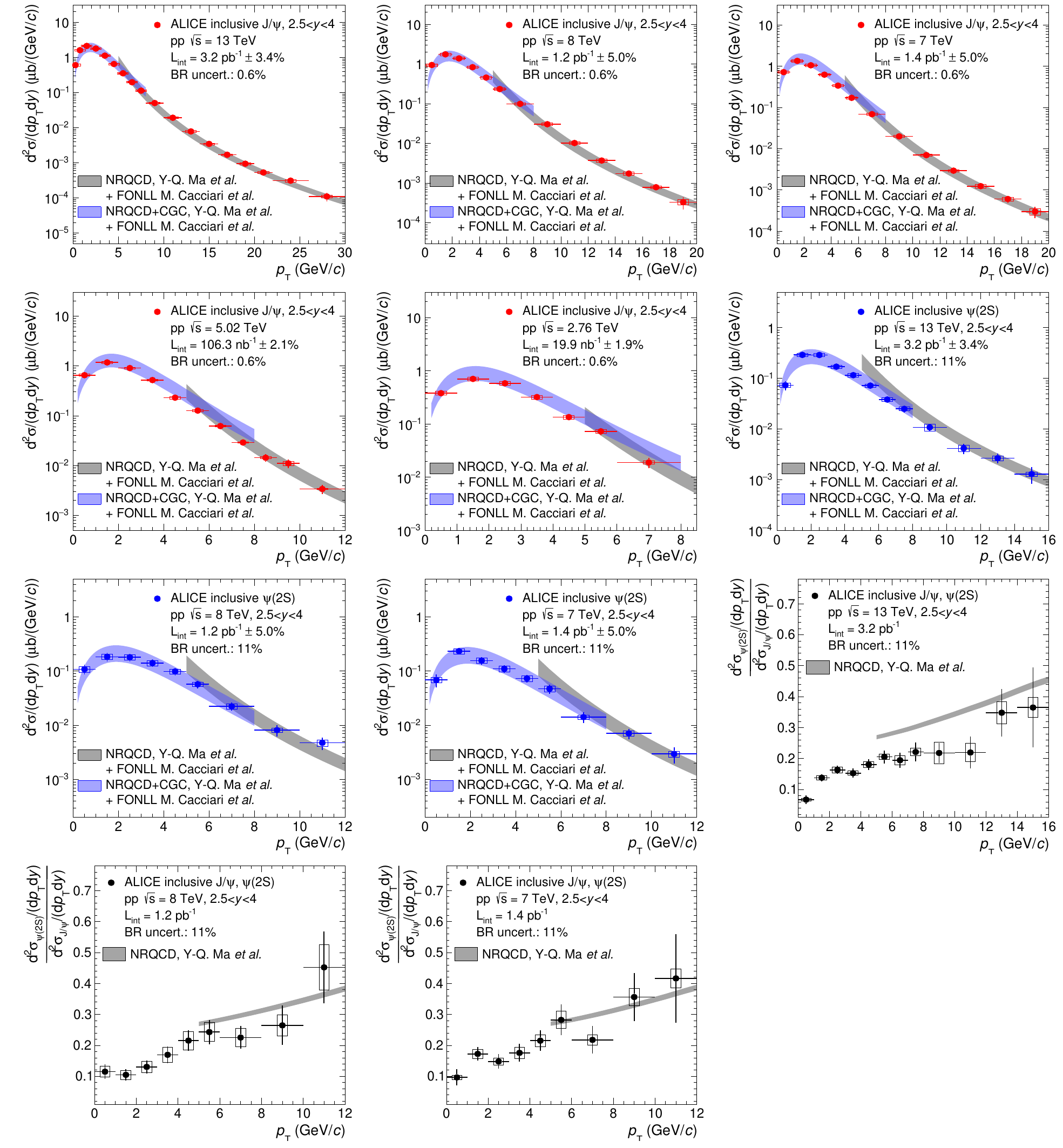}
\caption{\label{fig:theoryMa}(color online). Comparisons between ALICE $\jpsi$ and $\psip$ data and summed NRQCD and FONLL model calculations from~\cite{Ma:2010yw}, ~\cite{Ma:2014mri} and ~\cite{Cacciari:2012ny}. The first five panels correspond to inclusive $\jpsi$ production cross sections as a function of $\pt$ in pp collisions at $\sqrts=13$, $8$, $7$, $5.02$ and $2.76$~TeV (red), the next three panels to inclusive $\psip$ cross sections as a function of $\pt$ at $\sqrts=13$, $8$ and $7$ TeV (blue) and the last three panels to $\psip$-to-$\jpsi$ cross section ratios as a function of $\pt$ at the same $\sqrts$ (black).}
\end{figure}

The same groups have also provided NRQCD calculations for inclusive $\jpsi$ production in pp collisions at $\sqrts=8$, $7$, $5.02$ and $2.76$~TeV, and for $\psip$ at $\sqrts=13$, $8$ and $7$~TeV. These calculations are compared to ALICE measurements in Fig.~\ref{fig:theoryMa}. 
Also shown in this figure are comparisons from the high-$\pt$ NLO NRQCD calculations to ALICE $\psip$-to-$\jpsi$ cross section ratios as a function of $\pt$. The motivation for showing this comparison of the cross section ratios is that many of the systematic uncertainties cancel both for the data (as discussed in Sec.~\ref{sec:results13TeV}) and for the theory. 

Except for the cross section ratios, in all other panels the same strategy as in Fig.~\ref{fig:PtMa13TeV} is applied and the non-prompt contribution to inclusive charmonium production is added to the model using FONLL before comparing to the data. The FONLL+NRQCD summation is not performed for $\psip$-to-$\jpsi$ cross section ratios due to the added complexity introduced by the estimation of the error cancellation between the models. Moreover, the impact of the non-prompt charmonium contribution on these ratios is expected to be small because it enters both the numerator and the denominator with a similar magnitude (according to FONLL) and largely cancels out. We note that similar high-$\pt$ NLO NRQCD calculations~\cite{Ma:2010jj} were already compared to ALICE $\jpsi$ and $\psip$ cross sections at $\sqrts=7$~TeV in~\cite{Abelev:2014qha}, albeit with a different strategy to account for the non-prompt charmonia.

\begin{figure}[t!]
\renewcommand{\arraystretch}{0.5}
\includegraphics[width=\textwidth]{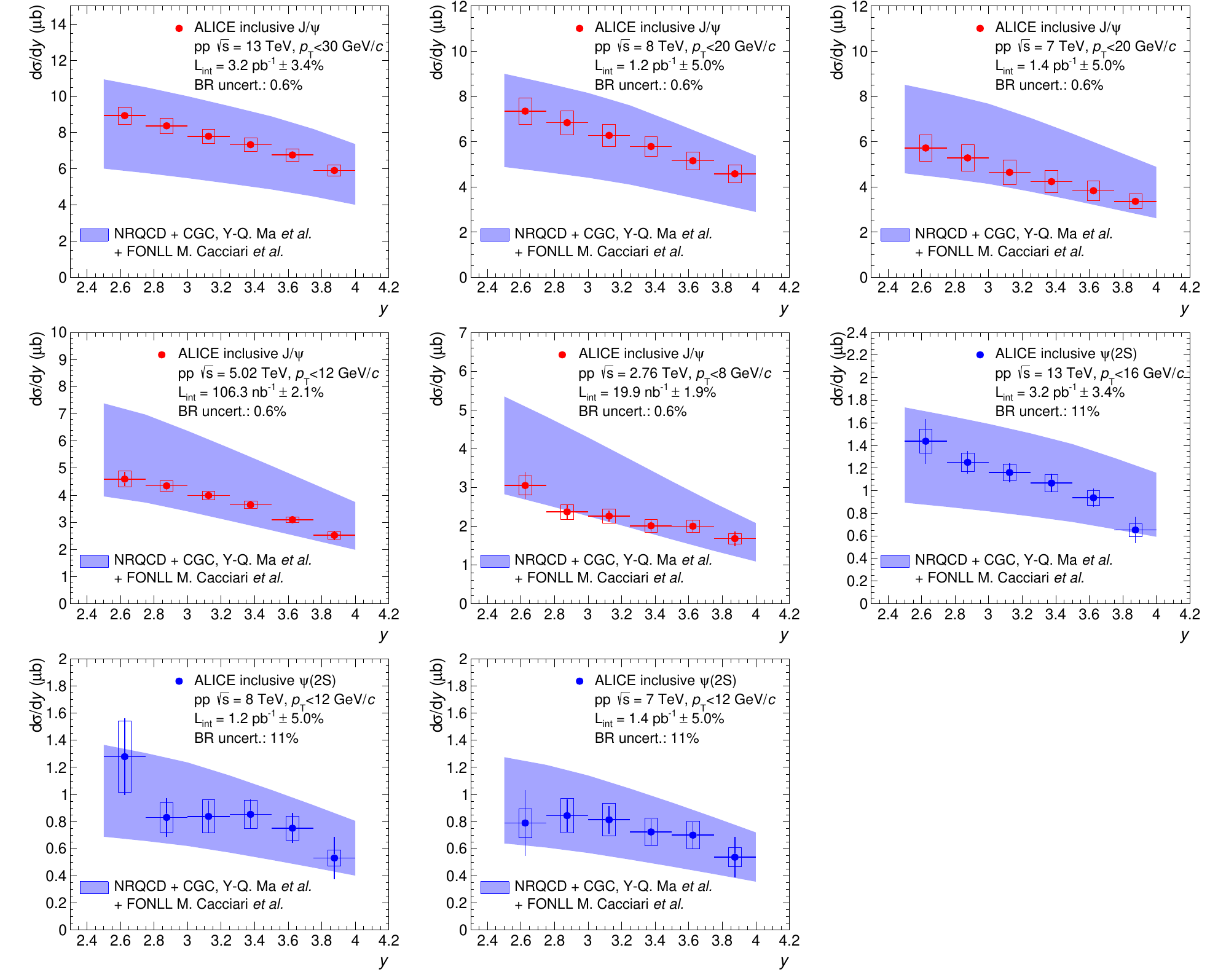}
\caption{\label{fig:theoryMa_y}(color online). Comparisons between ALICE $\jpsi$ and $\psip$ data and summed NRQCD and FONLL model calculations from~\cite{Ma:2014mri} and ~\cite{Cacciari:2012ny}. The first five panels correspond to inclusive $\jpsi$ production cross sections as a function of $y$ in pp collisions at $\sqrts=13$, $8$ and $7$, $5.02$ and $2.76$~TeV (red), while the next three panels to inclusive $\psip$ cross sections as a function of $y$ at $\sqrts=13$, $8$ and $7$ TeV (blue).}
\end{figure}

Since the NRQCD+CGC calculation from~\cite{Ma:2014mri} extends down to zero $\pt$, it can be integrated over $\pt$ and directly compared to ALICE $\pt$-integrated cross sections as a function of $y$. This calculation neglects the contribution from charmonium with $\pt>8$~GeV/$c$ to the total cross section, which anyway contributes by less than $3$\%. The results of this comparison as a function of $y$ are shown in Fig.~\ref{fig:theoryMa_y}.

Overall, a good agreement between the model and the data is observed for all measured cross sections, for both $\jpsi$ and $\psip$ as a function of either $\pt$ or $y$ and for all the collision energies considered. For $\psip$-to-$\jpsi$ cross section ratios as a function of $\pt$ however, the model tends to be slightly above the data especially at $\sqrts=13$~TeV. This tension appears mainly because of the error cancellation between the uncertainties on the $\jpsi$ and $\psip$ cross sections mentioned above.

\begin{figure}[t!]
\includegraphics[width=\textwidth]{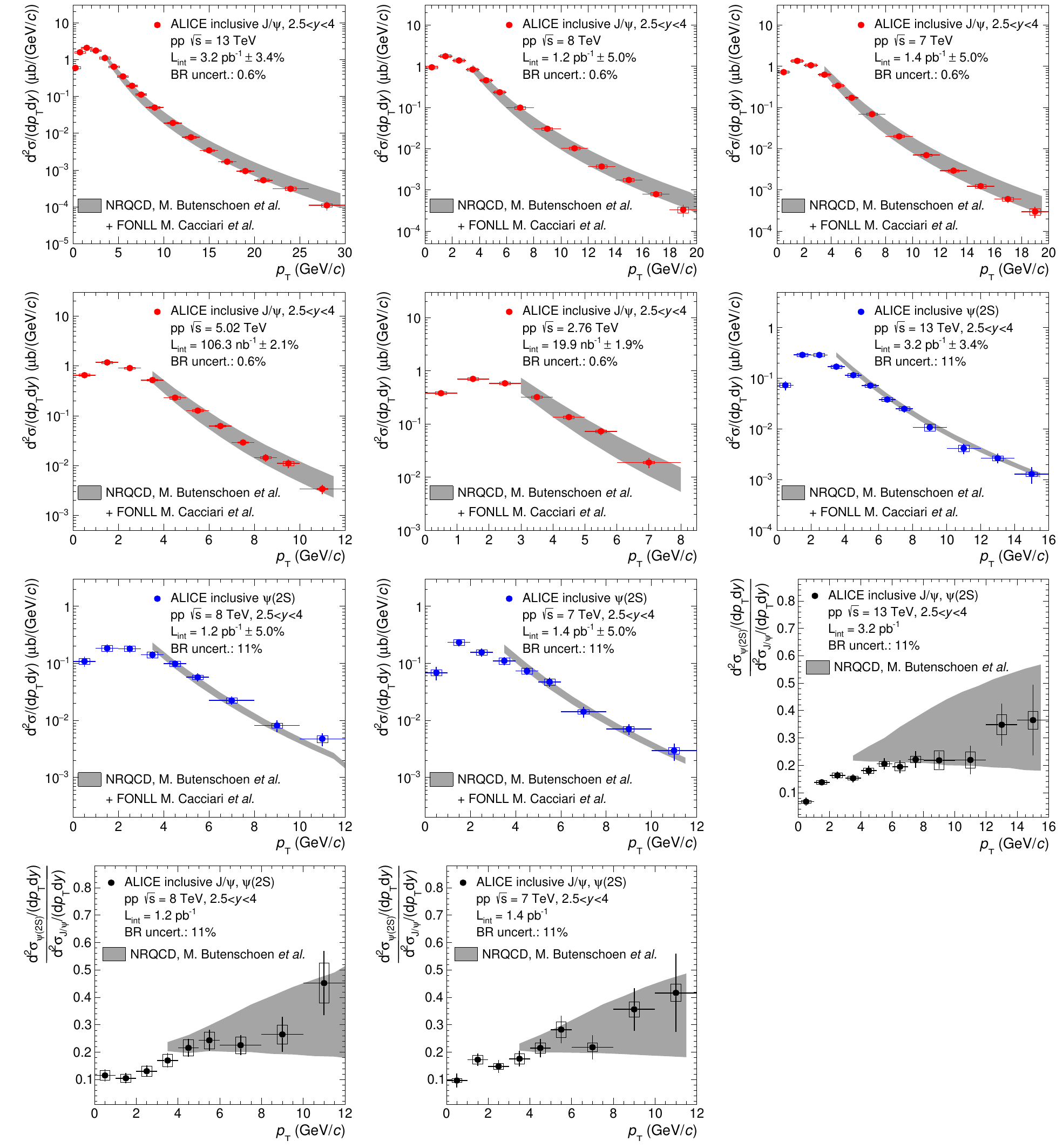}
\caption{(color online). Comparisons between ALICE $\jpsi$ and $\psip$ data and summed NRQCD and FONLL model calculations from~\cite{Butenschoen:2010rq} and ~\cite{Cacciari:2012ny}. The first five panels correspond to inclusive $\jpsi$ production cross sections as a function of $\pt$ in pp collisions at $\sqrts=13$, $8$, $7$, $5.02$ and $2.76$~TeV (red), the next three panels to inclusive $\psip$ cross sections as a function of $\pt$ at $\sqrts=13$, $8$ and $7$ TeV (blue) and the last three panels to $\psip$-to-$\jpsi$ cross section ratios as a function of $\pt$ at the same $\sqrts$ (black).}\label{fig:theorybutenschon}
\end{figure}

In Fig.~\ref{fig:theorybutenschon}, the ALICE measurements are compared to a second set of NLO NRQCD calculations from Butensch\"on and Kniehl~\cite{Butenschoen:2010rq}. In this case only high-$\pt$ calculations ($\pt>3$~GeV/$c$) are available. The ALICE $\pt$-integrated cross sections as a function of $y$ cannot be thus compared to the theory due to this $\pt$ cut. As was the case for the comparisons shown in Figs.~\ref{fig:theoryMa} and~\ref{fig:theoryMa_y}, FONLL is used to estimate the contribution from non-prompt charmonium production and added to the NRQCD calculation.

The two NLO NRQCD calculations from Butensch\"on and Kniehl (Fig.~\ref{fig:theorybutenschon}) and from Ma, Wang and Chao (Fig.~\ref{fig:theoryMa}) differ in the parametrization of the Long Distance Matrix Elements (LDME) used to calculate the color-octet contributions to the charmonium production cross section. The first calculation uses three matrix elements whereas the second uses only two linear combinations of these three elements. Other differences include: the data sets used to fit these matrix elements, the minimum $\pt$ above which the calculation is applicable and the way by which contributions from $\chi_c$ and $\psip$ decays to prompt $\jpsi$ production are accounted for. 

Although the agreement between the model and the data is of similar quality in Fig.~\ref{fig:theoryMa} and~\ref{fig:theorybutenschon}, some differences are visible. In particular, in Fig.~\ref{fig:theorybutenschon}, the calculation tends to overestimate the measured $\jpsi$ cross sections towards high-$\pt$ and the uncertainties are larger than in Fig.~\ref{fig:theoryMa}. The uncertainties on the $\psip$-to-$\jpsi$ cross section ratios are also significantly larger and consequently the agreement to the data is better. These observations are a consequence of the differences between the two calculations detailed above and in particular the fact that the fits of the LDME start at a lower $\pt$ and include a larger number of data sets in the second case.

\section{Conclusions}\label{section:conclusions}
The inclusive $\jpsi$ and $\psip$ differential cross sections as well as $\psip$-to-$\jpsi$ cross section ratios as a function of $\pt$ and $y$ have been measured in pp collisions at $\sqrts=5.02$ and $13$~TeV with the ALICE detector. Combined with similar measurements performed at $\sqrts=2.76$~\cite{Abelev:2012kr}, $7$~\cite{Abelev:2014qha} and $8$~TeV~\cite{Adam:2015rta}, these results constitute a stringent test for models of charmonium production and allow the study of quantities such as $\meanpt$, $\meanptsq$ and $\pt$-integrated $\dif\sigma/\dif y$ as a function of $\sqrts$.

The results at $\sqrts=$13~TeV significantly extend the $\pt$ reach for both charmonium states with respect to measurements performed by ALICE at lower energies, up to 30~GeV/$c$ for the $\jpsi$ and 16~GeV/$c$ for the $\psip$. When comparing the $\jpsi$ cross sections vs $\pt$ to measurements at lower $\sqrts$, a hardening of the spectra is observed with increasing collision energy. This is confirmed by measurements of the $\jpsi$ $\meanpt$ and $\meanptsq$, while a similar trend is observed for the $\psip$. Regarding inclusive $\psip$-to-$\jpsi$ cross section ratios, no $\sqrts$ dependence is observed within uncertainties. 

Comparisons of $\jpsi$ and $\psip$ cross sections and cross section ratios as a function of both $\pt$ and $y$ to NLO NRQCD and LO NRQCD+CGC prompt-charmonium calculations have been presented for all available collision energies. Concerning the $\jpsi$ cross section as a function of $\pt$, an excellent agreement is observed between data and theory, provided that the non-prompt contribution to the inclusive cross section is included using FONLL. This comparison indicates that for $\pt>15$~GeV/$c$, the non-prompt contribution can reach up to $50$\%. 
An overall good agreement is also observed for $\psip$ production and for the cross sections as a function of $\y$ albeit with larger uncertainties.

With the large contribution from non-prompt $\jpsi$ to the inclusive cross sections observed for high $\pt$ at $\sqrt{s}=13$~TeV, it is of relatively little interest to try to further extend the $\pt$ reach of the inclusive measurement for understanding charmonium production. This is as long as one is not capable of separating experimentally the prompt and the non-prompt contributions and relies on models instead. This separation will become possible in ALICE starting from 2021 with the addition of the Muon Forward Tracker~\cite{CERN-LHCC-2015-001}.

%%%%% acknowledgements
\newenvironment{acknowledgement}{\relax}{\relax}
\begin{acknowledgement}
\section*{Acknowledgements}

The ALICE Collaboration would like to thank Mathias Butensch\"on, Matteo Cacciari, Yan-Qing Ma and Ramona Vogt for providing the NRQCD, FONLL and CEM calculations used in this paper. 
% Version: 2017-01-20

The ALICE Collaboration would like to thank all its engineers and technicians for their invaluable contributions to the construction of the experiment and the CERN accelerator teams for the outstanding performance of the LHC complex.
The ALICE Collaboration gratefully acknowledges the resources and support provided by all Grid centres and the Worldwide LHC Computing Grid (WLCG) collaboration.
The ALICE Collaboration acknowledges the following funding agencies for their support in building and running the ALICE detector:
A. I. Alikhanyan National Science Laboratory (Yerevan Physics Institute) Foundation (ANSL), State Committee of Science and World Federation of Scientists (WFS), Armenia;
Austrian Academy of Sciences and Nationalstiftung f\"{u}r Forschung, Technologie und Entwicklung, Austria;
Ministry of Communications and High Technologies, National Nuclear Research Center, Azerbaijan;
Conselho Nacional de Desenvolvimento Cient\'{\i}fico e Tecnol\'{o}gico (CNPq), Universidade Federal do Rio Grande do Sul (UFRGS), Financiadora de Estudos e Projetos (Finep) and Funda\c{c}\~{a}o de Amparo \`{a} Pesquisa do Estado de S\~{a}o Paulo (FAPESP), Brazil;
Ministry of Science \& Technology of China (MSTC), National Natural Science Foundation of China (NSFC) and Ministry of Education of China (MOEC) , China;
Ministry of Science, Education and Sport and Croatian Science Foundation, Croatia;
Ministry of Education, Youth and Sports of the Czech Republic, Czech Republic;
The Danish Council for Independent Research | Natural Sciences, the Carlsberg Foundation and Danish National Research Foundation (DNRF), Denmark;
Helsinki Institute of Physics (HIP), Finland;
Commissariat \`{a} l'Energie Atomique (CEA) and Institut National de Physique Nucl\'{e}aire et de Physique des Particules (IN2P3) and Centre National de la Recherche Scientifique (CNRS), France;
Bundesministerium f\"{u}r Bildung, Wissenschaft, Forschung und Technologie (BMBF) and GSI Helmholtzzentrum f\"{u}r Schwerionenforschung GmbH, Germany;
Ministry of Education, Research and Religious Affairs, Greece;
National Research, Development and Innovation Office, Hungary;
Department of Atomic Energy Government of India (DAE) and Council of Scientific and Industrial Research (CSIR), New Delhi, India;
Indonesian Institute of Science, Indonesia;
Centro Fermi - Museo Storico della Fisica e Centro Studi e Ricerche Enrico Fermi and Istituto Nazionale di Fisica Nucleare (INFN), Italy;
Institute for Innovative Science and Technology , Nagasaki Institute of Applied Science (IIST), Japan Society for the Promotion of Science (JSPS) KAKENHI and Japanese Ministry of Education, Culture, Sports, Science and Technology (MEXT), Japan;
Consejo Nacional de Ciencia (CONACYT) y Tecnolog\'{i}a, through Fondo de Cooperaci\'{o}n Internacional en Ciencia y Tecnolog\'{i}a (FONCICYT) and Direcci\'{o}n General de Asuntos del Personal Academico (DGAPA), Mexico;
Nationaal instituut voor subatomaire fysica (Nikhef), Netherlands;
The Research Council of Norway, Norway;
Commission on Science and Technology for Sustainable Development in the South (COMSATS), Pakistan;
Pontificia Universidad Cat\'{o}lica del Per\'{u}, Peru;
Ministry of Science and Higher Education and National Science Centre, Poland;
Korea Institute of Science and Technology Information and National Research Foundation of Korea (NRF), Republic of Korea;
Ministry of Education and Scientific Research, Institute of Atomic Physics and Romanian National Agency for Science, Technology and Innovation, Romania;
Joint Institute for Nuclear Research (JINR), Ministry of Education and Science of the Russian Federation and National Research Centre Kurchatov Institute, Russia;
Ministry of Education, Science, Research and Sport of the Slovak Republic, Slovakia;
National Research Foundation of South Africa, South Africa;
Centro de Aplicaciones Tecnol\'{o}gicas y Desarrollo Nuclear (CEADEN), Cubaenerg\'{\i}a, Cuba, Ministerio de Ciencia e Innovacion and Centro de Investigaciones Energ\'{e}ticas, Medioambientales y Tecnol\'{o}gicas (CIEMAT), Spain;
Swedish Research Council (VR) and Knut \& Alice Wallenberg Foundation (KAW), Sweden;
European Organization for Nuclear Research, Switzerland;
National Science and Technology Development Agency (NSDTA), Suranaree University of Technology (SUT) and Office of the Higher Education Commission under NRU project of Thailand, Thailand;
Turkish Atomic Energy Agency (TAEK), Turkey;
National Academy of  Sciences of Ukraine, Ukraine;
Science and Technology Facilities Council (STFC), United Kingdom;
National Science Foundation of the United States of America (NSF) and United States Department of Energy, Office of Nuclear Physics (DOE NP), United States of America.

\end{acknowledgement}

%%%%%%%% Bibliography (In case of using bibtex generate the bbl requested by arXiv)
\bibliographystyle{utphys}   % Remember we use title in the biblio
\bibliography{bibliography}
%\input {bibliography.tex}  

%%%%%%%%% appendix with author list
\newpage
\appendix
\section{The ALICE Collaboration}
\label{app:collab}

% Collaboration: CERN-LHC-ALICE
% Generation Date is 2017-Jan-20

% How to use:
%%%%%%%%% appendix with author list
%\appendix
%\section{The ALICE Collaboration}
%\label{app:collab}
%\input{authors-list.tex}  %%%%%%% get the latest version before submitting

\begingroup
\small
\begin{flushleft}
S.~Acharya$^\textrm{\scriptsize 139}$,
D.~Adamov\'{a}$^\textrm{\scriptsize 87}$,
M.M.~Aggarwal$^\textrm{\scriptsize 91}$,
G.~Aglieri Rinella$^\textrm{\scriptsize 34}$,
M.~Agnello$^\textrm{\scriptsize 30}$,
N.~Agrawal$^\textrm{\scriptsize 47}$,
Z.~Ahammed$^\textrm{\scriptsize 139}$,
N.~Ahmad$^\textrm{\scriptsize 17}$,
S.U.~Ahn$^\textrm{\scriptsize 69}$,
S.~Aiola$^\textrm{\scriptsize 143}$
% ...
\renewcommand\labelenumi{\textsuperscript{\theenumi}~}

\section*{Affiliation notes}
\renewcommand\theenumi{\roman{enumi}}
\begin{Authlist}
\item \Adef{0}Deceased
\item \Adef{idp1805344}{Also at: Dipartimento DET del Politecnico di Torino, Turin, Italy}
\item \Adef{idp1824736}{Also at: Georgia State University, Atlanta, Georgia, United States}
\item \Adef{idp3259648}{Also at: Also at Department of Applied Physics, Aligarh Muslim University, Aligarh, India}
\item \Adef{idp4031136}{Also at: M.V. Lomonosov Moscow State University, D.V. Skobeltsyn Institute of Nuclear, Physics, Moscow, Russia}
\end{Authlist}

\section*{Collaboration Institutes}
\renewcommand\theenumi{\arabic{enumi}~}

$^{1}$A.I. Alikhanyan National Science Laboratory (Yerevan Physics Institute) Foundation, Yerevan, Armenia
\\
$^{2}$Benem\'{e}rita Universidad Aut\'{o}noma de Puebla, Puebla, Mexico
\\
$^{3}$Bogolyubov Institute for Theoretical Physics, Kiev, Ukraine
\\
$^{4}$Bose Institute, Department of Physics 
and Centre for Astroparticle Physics and Space Science (CAPSS), Kolkata, India
\\
$^{5}$Budker Institute for Nuclear Physics, Novosibirsk, Russia
\\
$^{6}$California Polytechnic State University, San Luis Obispo, California, United States
\\
$^{7}$Central China Normal University, Wuhan, China
\\
$^{8}$Centre de Calcul de l'IN2P3, Villeurbanne, Lyon, France
\\
$^{9}$Centro de Aplicaciones Tecnol\'{o}gicas y Desarrollo Nuclear (CEADEN), Havana, Cuba
\\
$^{10}$Centro de Investigaciones Energ\'{e}ticas Medioambientales y Tecnol\'{o}gicas (CIEMAT), Madrid, Spain
\\
$^{11}$Centro de Investigaci\'{o}n y de Estudios Avanzados (CINVESTAV), Mexico City and M\'{e}rida, Mexico
\\
$^{12}$Centro Fermi - Museo Storico della Fisica e Centro Studi e Ricerche ``Enrico Fermi', Rome, Italy
\\
$^{13}$Chicago State University, Chicago, Illinois, United States
\\
$^{14}$China Institute of Atomic Energy, Beijing, China
\\
$^{15}$COMSATS Institute of Information Technology (CIIT), Islamabad, Pakistan
\\
$^{16}$Departamento de F\'{\i}sica de Part\'{\i}culas and IGFAE, Universidad de Santiago de Compostela, Santiago de Compostela, Spain
\\
$^{17}$Department of Physics, Aligarh Muslim University, Aligarh, India
\\
$^{18}$Department of Physics, Ohio State University, Columbus, Ohio, United States
\\
$^{19}$Department of Physics, Sejong University, Seoul, South Korea
\\
$^{20}$Department of Physics, University of Oslo, Oslo, Norway
\\
$^{21}$Department of Physics and Technology, University of Bergen, Bergen, Norway
\\
$^{22}$Dipartimento di Fisica dell'Universit\`{a} 'La Sapienza'
and Sezione INFN, Rome, Italy
\\
$^{23}$Dipartimento di Fisica dell'Universit\`{a}
and Sezione INFN, Cagliari, Italy
\\
$^{24}$Dipartimento di Fisica dell'Universit\`{a}
and Sezione INFN, Trieste, Italy
\\
$^{25}$Dipartimento di Fisica dell'Universit\`{a}
and Sezione INFN, Turin, Italy
\\
$^{26}$Dipartimento di Fisica e Astronomia dell'Universit\`{a}
and Sezione INFN, Bologna, Italy
\\
$^{27}$Dipartimento di Fisica e Astronomia dell'Universit\`{a}
and Sezione INFN, Catania, Italy
\\
$^{28}$Dipartimento di Fisica e Astronomia dell'Universit\`{a}
and Sezione INFN, Padova, Italy
\\
$^{29}$Dipartimento di Fisica `E.R.~Caianiello' dell'Universit\`{a}
and Gruppo Collegato INFN, Salerno, Italy
\\
$^{30}$Dipartimento DISAT del Politecnico and Sezione INFN, Turin, Italy
\\
$^{31}$Dipartimento di Scienze e Innovazione Tecnologica dell'Universit\`{a} del Piemonte Orientale and INFN Sezione di Torino, Alessandria, Italy
\\
$^{32}$Dipartimento Interateneo di Fisica `M.~Merlin'
and Sezione INFN, Bari, Italy
\\
$^{33}$Division of Experimental High Energy Physics, University of Lund, Lund, Sweden
\\
$^{34}$European Organization for Nuclear Research (CERN), Geneva, Switzerland
\\
$^{35}$Excellence Cluster Universe, Technische Universit\"{a}t M\"{u}nchen, Munich, Germany
\\
$^{36}$Faculty of Engineering, Bergen University College, Bergen, Norway
\\
$^{37}$Faculty of Mathematics, Physics and Informatics, Comenius University, Bratislava, Slovakia
\\
$^{38}$Faculty of Nuclear Sciences and Physical Engineering, Czech Technical University in Prague, Prague, Czech Republic
\\
$^{39}$Faculty of Science, P.J.~\v{S}af\'{a}rik University, Ko\v{s}ice, Slovakia
\\
$^{40}$Faculty of Technology, Buskerud and Vestfold University College, Tonsberg, Norway
\\
$^{41}$Frankfurt Institute for Advanced Studies, Johann Wolfgang Goethe-Universit\"{a}t Frankfurt, Frankfurt, Germany
\\
$^{42}$Gangneung-Wonju National University, Gangneung, South Korea
\\
$^{43}$Gauhati University, Department of Physics, Guwahati, India
\\
$^{44}$Helmholtz-Institut f\"{u}r Strahlen- und Kernphysik, Rheinische Friedrich-Wilhelms-Universit\"{a}t Bonn, Bonn, Germany
\\
$^{45}$Helsinki Institute of Physics (HIP), Helsinki, Finland
\\
$^{46}$Hiroshima University, Hiroshima, Japan
\\
$^{47}$Indian Institute of Technology Bombay (IIT), Mumbai, India
\\
$^{48}$Indian Institute of Technology Indore, Indore, India
\\
$^{49}$Indonesian Institute of Sciences, Jakarta, Indonesia
\\
$^{50}$Inha University, Incheon, South Korea
\\
$^{51}$Institut de Physique Nucl\'eaire d'Orsay (IPNO), Universit\'e Paris-Sud, CNRS-IN2P3, Orsay, France
\\
$^{52}$Institute for Nuclear Research, Academy of Sciences, Moscow, Russia
\\
$^{53}$Institute for Subatomic Physics of Utrecht University, Utrecht, Netherlands
\\
$^{54}$Institute for Theoretical and Experimental Physics, Moscow, Russia
\\
$^{55}$Institute of Experimental Physics, Slovak Academy of Sciences, Ko\v{s}ice, Slovakia
\\
$^{56}$Institute of Physics, Academy of Sciences of the Czech Republic, Prague, Czech Republic
\\
$^{57}$Institute of Physics, Bhubaneswar, India
\\
$^{58}$Institute of Space Science (ISS), Bucharest, Romania
\\
$^{59}$Institut f\"{u}r Informatik, Johann Wolfgang Goethe-Universit\"{a}t Frankfurt, Frankfurt, Germany
\\
$^{60}$Institut f\"{u}r Kernphysik, Johann Wolfgang Goethe-Universit\"{a}t Frankfurt, Frankfurt, Germany
\\
$^{61}$Institut f\"{u}r Kernphysik, Westf\"{a}lische Wilhelms-Universit\"{a}t M\"{u}nster, M\"{u}nster, Germany
\\
$^{62}$Instituto de Ciencias Nucleares, Universidad Nacional Aut\'{o}noma de M\'{e}xico, Mexico City, Mexico
\\
$^{63}$Instituto de F\'{i}sica, Universidade Federal do Rio Grande do Sul (UFRGS), Porto Alegre, Brazil
\\
$^{64}$Instituto de F\'{\i}sica, Universidad Nacional Aut\'{o}noma de M\'{e}xico, Mexico City, Mexico
\\
$^{65}$IRFU, CEA, Universit\'{e} Paris-Saclay, F-91191 Gif-sur-Yvette, France, Saclay, France
\\
$^{66}$iThemba LABS, National Research Foundation, Somerset West, South Africa
\\
$^{67}$Joint Institute for Nuclear Research (JINR), Dubna, Russia
\\
$^{68}$Konkuk University, Seoul, South Korea
\\
$^{69}$Korea Institute of Science and Technology Information, Daejeon, South Korea
\\
$^{70}$KTO Karatay University, Konya, Turkey
\\
$^{71}$Laboratoire de Physique Corpusculaire (LPC), Clermont Universit\'{e}, Universit\'{e} Blaise Pascal, CNRS--IN2P3, Clermont-Ferrand, France
\\
$^{72}$Laboratoire de Physique Subatomique et de Cosmologie, Universit\'{e} Grenoble-Alpes, CNRS-IN2P3, Grenoble, France
\\
$^{73}$Laboratori Nazionali di Frascati, INFN, Frascati, Italy
\\
$^{74}$Laboratori Nazionali di Legnaro, INFN, Legnaro, Italy
\\
$^{75}$Lawrence Berkeley National Laboratory, Berkeley, California, United States
\\
$^{76}$Moscow Engineering Physics Institute, Moscow, Russia
\\
$^{77}$Nagasaki Institute of Applied Science, Nagasaki, Japan
\\
$^{78}$National and Kapodistrian University of Athens, Physics Department, Athens, Greece, Athens, Greece
\\
$^{79}$National Centre for Nuclear Studies, Warsaw, Poland
\\
$^{80}$National Institute for Physics and Nuclear Engineering, Bucharest, Romania
\\
$^{81}$National Institute of Science Education and Research, Bhubaneswar, India
\\
$^{82}$National Nuclear Research Center, Baku, Azerbaijan
\\
$^{83}$National Research Centre Kurchatov Institute, Moscow, Russia
\\
$^{84}$Niels Bohr Institute, University of Copenhagen, Copenhagen, Denmark
\\
$^{85}$Nikhef, Nationaal instituut voor subatomaire fysica, Amsterdam, Netherlands
\\
$^{86}$Nuclear Physics Group, STFC Daresbury Laboratory, Daresbury, United Kingdom
\\
$^{87}$Nuclear Physics Institute, Academy of Sciences of the Czech Republic, \v{R}e\v{z} u Prahy, Czech Republic
\\
$^{88}$Oak Ridge National Laboratory, Oak Ridge, Tennessee, United States
\\
$^{89}$Petersburg Nuclear Physics Institute, Gatchina, Russia
\\
$^{90}$Physics Department, Creighton University, Omaha, Nebraska, United States
\\
$^{91}$Physics Department, Panjab University, Chandigarh, India
\\
$^{92}$Physics Department, University of Cape Town, Cape Town, South Africa
\\
$^{93}$Physics Department, University of Jammu, Jammu, India
\\
$^{94}$Physics Department, University of Rajasthan, Jaipur, India
\\
$^{95}$Physikalisches Institut, Eberhard Karls Universit\"{a}t T\"{u}bingen, T\"{u}bingen, Germany
\\
$^{96}$Physikalisches Institut, Ruprecht-Karls-Universit\"{a}t Heidelberg, Heidelberg, Germany
\\
$^{97}$Physik Department, Technische Universit\"{a}t M\"{u}nchen, Munich, Germany
\\
$^{98}$Purdue University, West Lafayette, Indiana, United States
\\
$^{99}$Pusan National University, Pusan, South Korea
\\
$^{100}$Research Division and ExtreMe Matter Institute EMMI, GSI Helmholtzzentrum f\"ur Schwerionenforschung GmbH, Darmstadt, Germany
\\
$^{101}$Rudjer Bo\v{s}kovi\'{c} Institute, Zagreb, Croatia
\\
$^{102}$Russian Federal Nuclear Center (VNIIEF), Sarov, Russia
\\
$^{103}$Saha Institute of Nuclear Physics, Kolkata, India
\\
$^{104}$School of Physics and Astronomy, University of Birmingham, Birmingham, United Kingdom
\\
$^{105}$Secci\'{o}n F\'{\i}sica, Departamento de Ciencias, Pontificia Universidad Cat\'{o}lica del Per\'{u}, Lima, Peru
\\
$^{106}$Sezione INFN, Bari, Italy
\\
$^{107}$Sezione INFN, Bologna, Italy
\\
$^{108}$Sezione INFN, Cagliari, Italy
\\
$^{109}$Sezione INFN, Catania, Italy
\\
$^{110}$Sezione INFN, Padova, Italy
\\
$^{111}$Sezione INFN, Rome, Italy
\\
$^{112}$Sezione INFN, Trieste, Italy
\\
$^{113}$Sezione INFN, Turin, Italy
\\
$^{114}$SSC IHEP of NRC Kurchatov institute, Protvino, Russia
\\
$^{115}$Stefan Meyer Institut f\"{u}r Subatomare Physik (SMI), Vienna, Austria
\\
$^{116}$SUBATECH, IMT Atlantique, Universit\'{e} de Nantes, CNRS-IN2P3, Nantes, France
\\
$^{117}$Suranaree University of Technology, Nakhon Ratchasima, Thailand
\\
$^{118}$Technical University of Ko\v{s}ice, Ko\v{s}ice, Slovakia
\\
$^{119}$Technical University of Split FESB, Split, Croatia
\\
$^{120}$The Henryk Niewodniczanski Institute of Nuclear Physics, Polish Academy of Sciences, Cracow, Poland
\\
$^{121}$The University of Texas at Austin, Physics Department, Austin, Texas, United States
\\
$^{122}$Universidad Aut\'{o}noma de Sinaloa, Culiac\'{a}n, Mexico
\\
$^{123}$Universidade de S\~{a}o Paulo (USP), S\~{a}o Paulo, Brazil
\\
$^{124}$Universidade Estadual de Campinas (UNICAMP), Campinas, Brazil
\\
$^{125}$Universidade Federal do ABC, Santo Andre, Brazil
\\
$^{126}$University of Houston, Houston, Texas, United States
\\
$^{127}$University of Jyv\"{a}skyl\"{a}, Jyv\"{a}skyl\"{a}, Finland
\\
$^{128}$University of Liverpool, Liverpool, United Kingdom
\\
$^{129}$University of Tennessee, Knoxville, Tennessee, United States
\\
$^{130}$University of the Witwatersrand, Johannesburg, South Africa
\\
$^{131}$University of Tokyo, Tokyo, Japan
\\
$^{132}$University of Tsukuba, Tsukuba, Japan
\\
$^{133}$University of Zagreb, Zagreb, Croatia
\\
$^{134}$Universit\'{e} de Lyon, Universit\'{e} Lyon 1, CNRS/IN2P3, IPN-Lyon, Villeurbanne, Lyon, France
\\
$^{135}$Universit\'{e} de Strasbourg, CNRS, IPHC UMR 7178, F-67000 Strasbourg, France, Strasbourg, France
\\
$^{136}$Universit\`{a} degli Studi di Pavia, Pavia, Italy
\\
$^{137}$Universit\`{a} di Brescia, Brescia, Italy
\\
$^{138}$V.~Fock Institute for Physics, St. Petersburg State University, St. Petersburg, Russia
\\
$^{139}$Variable Energy Cyclotron Centre, Kolkata, India
\\
$^{140}$Warsaw University of Technology, Warsaw, Poland
\\
$^{141}$Wayne State University, Detroit, Michigan, United States
\\
$^{142}$Wigner Research Centre for Physics, Hungarian Academy of Sciences, Budapest, Hungary
\\
$^{143}$Yale University, New Haven, Connecticut, United States
\\
$^{144}$Yonsei University, Seoul, South Korea
\\
$^{145}$Zentrum f\"{u}r Technologietransfer und Telekommunikation (ZTT), Fachhochschule Worms, Worms, Germany
\endgroup

  %%%%%%% done by webmaster team
\end{document}